\documentclass[a4paper,11pt]{article} 
\usepackage[margin=0.75in]{geometry} 
\usepackage{amssymb}
\usepackage{amsmath,epsfig, epsf, epic, graphicx} 
\usepackage{fancybox, lscape} 
\usepackage{bm}
\usepackage{natbib}
\usepackage{graphicx}
\DeclareGraphicsExtensions{.eps,.pdf,.jpeg,.png}
\usepackage{caption}
\usepackage[margin=2cm]{caption}
\usepackage{setspace}
\usepackage{cleveref}
\usepackage{xcolor}
\usepackage{array}
\usepackage{wrapfig}
\usepackage{multirow}
\usepackage{url}
\usepackage{tabularx}
\allowdisplaybreaks

\usepackage{floatrow}
\usepackage{subfig}

\title{Multiple exposure distributed lag models with variable selection}
\author{Joseph Antonelli, Ander Wilson, and Brent Coull}

\begin{document}

\maketitle

\begin{abstract}
{Distributed lag models are useful in environmental epidemiology as they allow the user to investigate critical windows of exposure, defined as the time period during which exposure to a pollutant adversely affects health outcomes. Recent studies have focused on estimating the health effects of a large number of environmental exposures, or an environmental mixture, on health outcomes. In such settings, it is important to understand which environmental exposures affect a particular outcome, while acknowledging the possibility that different exposures have different critical windows.   Further, in the studies of environmental mixtures, it is important to identify interactions among exposures, and to account for the fact that this interaction may occur between two exposures having different critical windows. Exposure to one exposure early in time could cause an individual to be more or less susceptible to another exposure later in time.   We propose a Bayesian model to estimate the temporal effects of a large number of exposures on an outcome. We use spike-and-slab priors and semiparametric distributed lag curves to identify important exposures and exposure interactions, and discuss extensions with improved power to detect harmful exposures. We then apply these methods to estimate the effects of exposure to multiple air pollutants during pregnancy on birthweight from vital records in Colorado.}
\end{abstract}

\section{Introduction}

Humans are continuously exposed to a complex mixture of environmental chemicals. Starting with conception, these exposures can perturb the developmental process and alter children's health outcomes. Recent research has focused on identifying critical windows of susceptibility, which are times during the developmental process when an exposure can alter a future phenotype \citep{Wright2017}. Identifying critical windows requires exposure data that are observed repeatedly during the developmental period of interest, with many studies focusing on gestation. Numerous studies have estimated critical windows and the exposure-time-response relationship for exposure to a single environmental chemical during pregnancy \citep{Lee2017,Bose2017a,Chiu2017}. However, there remains a gap with respect to statistical methods that can identify critical windows to a mixture of multiple environmental pollutants.

A popular approach to estimate the association between maternal exposure during pregnancy and a birth outcome is the distributed lag model (DLM) \citep{schwartz2000distributed,zanobetti2000generalized,Darrow2010a}. In a DLM, an outcome is regressed on repeated measures of exposures over a preceding time period. Compared to using exposures averaged over a pre-specified time window, such as trimester average exposures, a DLM has been shown to reduce bias and improve estimation of critical windows \citep{Wilson2017}. The recent statistical literature has extended DLM methods in a number of directions, such as estimation of nonlinear exposure-time-response functions \citep{gasparrini2014modeling,gasparrini2017penalized, Mork2021TreedModels}, spatially-varying distributed lag functions  \citep{Warren2012,warren2020spatially}, subgroup specific distributed lag functions \citep{wilson2017bayesian}, and incorporation of variable selection procedures that select time periods in a critical window \citep{Warren2019}.

In this manuscript we focus on methods for the analysis of the health effects of a mixture of multiple environmental chemicals that are measured longitudinally. We have three primary objectives. The first is to select the mixture components and interactions that are associated with the outcome. When selecting mixture components and interactions, we are particularly focused on maximizing the power to detect components associated with the outcome while minimizing the false discovery rate. The second 
is to estimate the exposure-time-response relationship which includes both main effects and pairwise interactions of the mixture. Interactions between repeated measures of exposure may include interactions between two exposures at the same time point or time-lagged interactions. The third objective is to identify critical windows for each of the exposures and their interactions. 

There is a large body of research on statistical methods for mixtures that are observed at a single time point. For reviews see \cite{Taylor2016b}, \cite{Davalos2017} and \cite{Gibson2019}. Fewer approaches have examined repeated measures of exposure to a  mixture. Bivariate distributed lag surfaces were first introduced in \cite{muggeo2007bivariate} and \cite{chen2019distributed} explored different approaches to modeling bivariate distributed lag surfaces. However, these approaches only consider an interaction between two components. For mixtures of more than two exposures, \cite{liu2017lagged} proposed a lagged Bayesian kernel machine regression model for a mixture observed at a small number of time points, such as trimesters. \cite{wilson2020kernel} proposed Bayesian kernel machine regression distributed lag models for repeated measures of a mixture observed at high temporal resolution. However, this approach is computationally demanding and can currently only be fit to small or moderately-sized datasets. \cite{Bello2017} proposed an additive DLM model using the weighted quantile sum (WQS) framework \citep{Carrico2014} that only incorporates main effects. \cite{Mork2021EstimatingPairs} proposed a flexible regression tree approach for mixtures of time varying exposures. \cite{warren2021critical} proposed an extension of the critical window variable selection \citep{Warren2019} for mixtures observed at repeated measures. Because the approach focuses on selecting critical windows, the approach allows for main effects and interactions among exposures at a single time point but does not allow for time-lagged interactions or exposure selection.

In this paper, we adopt spike-and-slab prior distributions \citep{mitchell1988bayesian} within the distributed lag framework to identify important exposures of the mixture and interactions among exposures that are associated with the outcome. This extends the literature on distributed lag modeling for environmental mixtures in a number of ways. Our approach allows for pairwise interactions among a large number of environmental exposures, while also allowing for time-lagged interactions instead of assuming they occur at a single point in time. Further, we use a novel approach to hyperparameter selection that increases the power to detect important mixture effects without greatly increasing the number of false discoveries. This is crucially important because the effects of environmental mixtures are frequently weak to moderate, and existing approaches may be underpowered to detect these signals. 

\section{Modeling framework}

Throughout we assume the observed data are independent and identically distributed replicates of $\boldsymbol{\mathcal{D}}_i = (Y_i, \boldsymbol{X}_{i1}, \boldsymbol{X}_{i2}, \dots, \boldsymbol{X}_{ip}, \boldsymbol{C}_i)$ for $i = 1, \dots, n$. $Y_i$ represents a scalar outcome. $\boldsymbol{X}_{ij}$ is a $T$-dimensional vector representing individual $i$'s exposure to pollutant $j$ across $T$ time points, and $\boldsymbol{C}_i$ is a $q$-dimensional vector of covariates that could confound the association between the exposures and outcome. We restrict attention to continuous outcomes as extensions to binary outcomes are straightforward.  To simplify notation, we drop the $i$ subscript for the rest of the manuscript.  The main goal of our methodology will be to estimate the relationship between the $p$ exposures and the outcome conditional on $\boldsymbol{C}$. We adopt the following distributed lag model framework for the mean of the outcome:
\begin{align}
	E(Y \vert \boldsymbol{X}, \boldsymbol{C}) = \beta_0 + \boldsymbol{C}^T \boldsymbol{\beta}_C + \underbrace{\sum_{j=1}^p \sum_{t=1}^T \eta_{jt} X_{jt}}_{\text{Main effect curves}} + \underbrace{\sum_{j_1=2}^p \sum_{j_2 < j_1} \sum_{t_1=1}^T \sum_{t_2=1}^T \eta_{j_1 j_2 t_1 t_2} X_{j_1 t_1} X_{j_2 t_2}}_{\text{Interaction surfaces}}.
\end{align}
\noindent  This leads to a highly parameterized model as there are $pT$ parameters $\eta_{jt}$ and $\binom{p}{2} T^2$ parameters $\eta_{j_1 j_2 t_1 t_2}$. The dimension of the parameter space can be reduced in a number of ways. The first is to impose smoothness conditions on the distributed lag curves across time within a particular exposure. To achieve this, we assume a functional form for the distributed lag curves, $\boldsymbol{\eta}_j=(\eta_{j1},\dots,\eta_{jT})^T$ for $j=1,\dots,p$, instead of allowing there to be $T$ distinct parameters for each main effect function. We also assume the interaction surface parameters $\eta_{j_1 j_2 t_1 t_2}$ vary smoothly with respect to $t_1$ and $t_2$: 
\begin{align}
	\eta_{jt} &= \sum_{k=1}^{K_j} f_{j k}(t) \beta_{j k} =  \boldsymbol{f}_j^T(t) \boldsymbol{\beta}_j \\
  \eta_{j_1 j_2 t_1 t_2} &= \sum_{k=1}^{K_{j_1 j_2}} f_{j_1 j_2 k}(t_1, t_2) \beta_{j_1 j_2 k} = \boldsymbol{f}_{j_1 j_2}^T (t_1, t_2) \boldsymbol{\beta}_{j_1 j_2},
\end{align}

\noindent where $f_{j k}(t)$ and $f_{j_1 j_2 k}(t_1, t_2)$ are one and two-dimensional basis functions of time that must be chosen {\em a priori}. This yields the following representation of the model:
\begin{align}
	E(Y \vert \boldsymbol{X}, \boldsymbol{C}) = \beta_0 + \boldsymbol{C}^T \boldsymbol{\beta}_C + \sum_{j=1}^p \sum_{t=1}^T \boldsymbol{f}_j^T(t) \boldsymbol{\beta}_j X_{jt} + \sum_{j_1=2}^p \sum_{j_2 < j_1} \sum_{t_1=1}^T \sum_{t_2=1}^T \boldsymbol{f}_{j_1 j_2}^T (t_1, t_2) \boldsymbol{\beta}_{j_1 j_2} X_{j_1 t_1} X_{j_2 t_2}. \label{eqn:MeanComponent}
\end{align}

\noindent Notice that for exposure $j$ we have reduced the number of parameters of the main effect function from $T$ to $K_{j}$. For the interaction between exposures $j_1$ and $j_2$, we have reduced the number of parameters from $T^2$ to $K_{j_1 j_2}$.  In the following sections, we discuss different approaches to choosing the basis functions and the number of basis functions to include, as well as additional dimension reduction approaches for the interactions. 

In fitting the model we also perform exposure selection, which is of scientific interest in environmental epidemiology. We achieve this goal using the model specification: 
\begin{align}
	Y \vert \boldsymbol{X}, \boldsymbol{C} &\sim \mathcal{N} \Big( E(Y \vert \boldsymbol{X}, \boldsymbol{C}), \sigma^2 \Big) \label{eqn:MainModel} \\
    \boldsymbol{\beta}_j \vert \gamma_j &\sim (1 - \gamma_j) \delta_{\boldsymbol{0}} + \gamma_j \mathcal{MVN}(\boldsymbol{0}, \sigma^2 \sigma_{M}^2 \boldsymbol{I}) \hspace{1.93cm} \gamma_j \vert \tau_{M_j} \sim \text{Bernoulli}(\tau_{M_j}) \\
    \boldsymbol{\beta}_{jk} \vert \gamma_{jk} &\sim (1 - \gamma_{jk}) \delta_{\boldsymbol{0}} + \gamma_{jk} \mathcal{MVN}(\boldsymbol{0}, \sigma^2 \sigma_{I}^2 \boldsymbol{I}) \hspace{1.69cm} \gamma_{jk} \vert \tau_{I_{jk}} \sim \text{Bernoulli}(\tau_{I_{jk}}). 
\end{align}
\noindent The prior distribution for the regression coefficients is a mixture between a point mass at $\boldsymbol{0}$ and a multivariate normal distribution. We introduce latent, binary variables $\gamma_j$ and $\gamma_{jk}$ which indicate whether a particular exposure or exposure pair have an important main effect or interaction surface, respectively. These can be interpreted as indicating if a particular distributed lag function is important or not, since $\gamma_j = 0$ implies that $\boldsymbol{\beta}_j = 0$ and the entire main effect for exposure $j$ is removed from the model. Analogous interpretations can be assigned to the $\gamma_{jk}$ variables for interactions. We assign a diffuse normal prior on $(\beta_0, \boldsymbol{\beta}_C)$, and an inverse gamma prior distribution for $\sigma^2$. The remaining parameters, given by $\sigma_M^2$, $\sigma_I^2$, $\tau_{M_j}$, and $ \tau_{I_{jk}}$  are particularly influential for the performance of the approach and we  discuss prior elicitation strategies for them in greater detail in the following sections. 

\subsection{Selection of basis functions}
\label{sec:Basis}
The model parameterizes the distributed lag parameters as a weighted sum of basis functions given by $\boldsymbol{f}_j(t)$ and $\boldsymbol{f}_{j_1 j_2} (t_1, t_2)$ for the main effect and interaction surfaces, respectively. Our goal is to keep the dimension of the parameter space relatively small, while retaining as much flexibility in the resulting surface effect estimates as possible. Further, we prefer fitting algorithms that are data-driven and do not force the user to select tuning parameters, such as the number of basis functions or the location of knots, both of which can impact the results. One approach that obviates the need for an analyst to choose the number of basis functions is based on principle components. Following the ideas proposed in \cite{wilson2017bayesian}, we let $\boldsymbol{X}_j$ represent the $n$ by $T$ matrix of exposure $j$ values measured over time and compute the sample covariance matrix defined by $\boldsymbol{X}_j^T \boldsymbol{X}_j$. The eigenvectors of this matrix can then be used as basis functions to parameterize the distributed lags curves. The choice of the number of basis functions can be automated by retaining only the eigenvectors that explain a pre-specified proportion of the variability in the data, such as 95\%. We use the smoothed covariance matrix proposed in \cite{xiao2016fast} and implemented in the R package \texttt{refund} \citep{refundpack} to obtain smooth basis functions, which we refer to as FPCA basis functions. Additionally we use a constant as one of the basis functions to allow the mean of the distributed lag curve to differ from zero.

For specification of the two-dimensional functions that  model the distributed lag interaction surfaces,  power becomes a pressing issue because interactions in environmental studies typically have small to moderate signals.  Therefore it is important to also reduce the number of parameters for these surfaces in order to increase power to detect an effect. Toward this goal,  we use tensor products of the one-dimensional basis functions from above, which yields $K_{j_1} K_{j_2}$ functions defined as $f_{j_1 j_2 k} (t_1, t_2) = f_{j_1 k_1}(t_1) f_{j_2 k_2}(t_2)$ for $k=1, \dots, K_{j_1} K_{j_2}$. By assuming a smooth surface for the interaction distributed lag surfaces with this tensor product structure, we reduce the dimension of the parameter space for each interaction surface from $T^2$ parameters to $K_{j_1} K_{j_2}$. Depending on the values of $K_{j_1}$ and $K_{j_2}$, this can still lead to a large number of parameters representing the distributed lag interaction surfaces. While this allows for flexibility in the modeling of the distributed lag surfaces, the power to detect important exposure effects can still be low. To reduce the parameter space further, we first write the component of $(\ref{eqn:MeanComponent})$ stemming from the interaction between exposure $j_1$ and $j_2$ as:
\begin{align*}
    \sum_{t_1=1}^T \sum_{t_2=1}^T  \boldsymbol{f}_{j_1 j_2}^T (t_1, t_2) \boldsymbol{\beta}_{j_1 j_2} X_{j_1 t_1} X_{j_2 t_2} = \boldsymbol{X}^*_{j_1, j_2} \boldsymbol{\beta}_{j_1 j_2}.
\end{align*}
Let $\boldsymbol{W}_{j_1, j_2}$ be a $K_{j_1} K_{j_2}$ by $K_{j_1} K_{j_2}$ matrix whose columns are the eigenvectors of ${\boldsymbol{X}^*}^T_{j_1, j_2} \boldsymbol{X}^*_{j_1, j_2}$ and $\boldsymbol{W}_{j_1, j_2}^R$ be a $K_{j_1} K_{j_2}$ by $R$ matrix representing only the first $R \leq K_{j_1} K_{j_2}$ columns of $\boldsymbol{W}$. We model the interaction surfaces using $
    \widetilde{\boldsymbol{X}}_{j_1, j_2} \widetilde{\boldsymbol{\beta}}_{j_1 j_2} = \boldsymbol{X}^*_{j_1, j_2} \boldsymbol{W}_{j_1, j_2}^R \widetilde{\boldsymbol{\beta}}_{j_1 j_2}.$
The parameter vector $\widetilde{\boldsymbol{\beta}}_{j_1 j_2}$ now contains only $R$ parameters. If $R$ is large enough, $\boldsymbol{X}^*_{j_1, j_2} \boldsymbol{\beta}_{j_1 j_2} \approx \widetilde{\boldsymbol{X}}_{j_1, j_2} \widetilde{\boldsymbol{\beta}}_{j_1 j_2}$ and we are able to adequately estimate this interaction, but with much fewer parameters and therefore increased power to detect an effect. The columns of $\boldsymbol{X}^*_{j_1, j_2}$ tend to be highly correlated and therefore substantial dimension reduction can be achieved with little loss of information. Note that for well-chosen $R$ this can improve power substantially, though it does come with a loss of information. Further, if $R$ is chosen to be too small, then too much information will be lost and the power to detect important signals can decrease. We choose $R$ such that a large portion of the variability is maintained from the original $\boldsymbol{X}^*_{j_1, j_2}$, such as 99\%. This leads to improved power and more efficient estimation without too much loss of information in the interaction surfaces.

\subsection{Relationship between prior parameters and power}

\begin{figure}[h]
\centering
	  \includegraphics[width=0.4\textwidth]{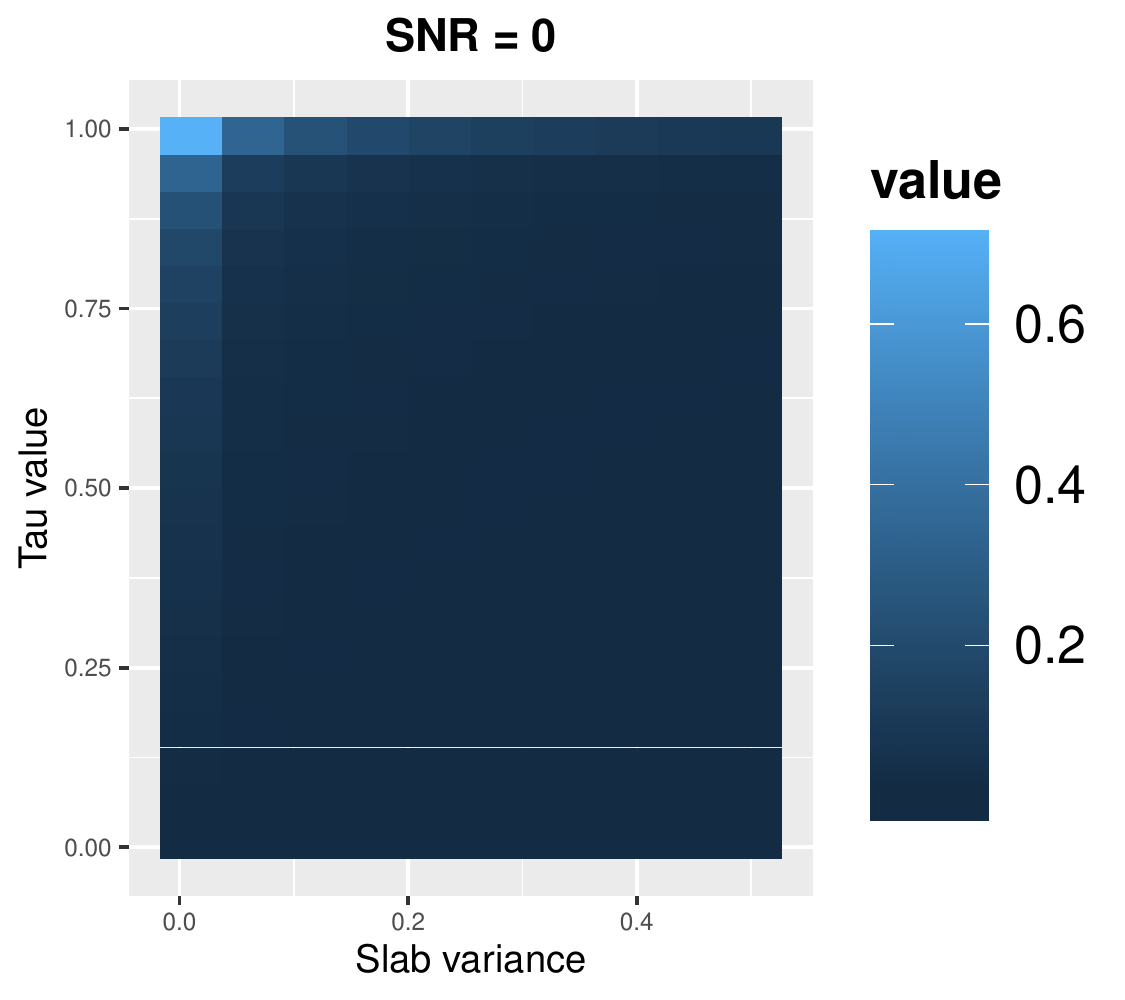}
      \includegraphics[width=0.4\textwidth]{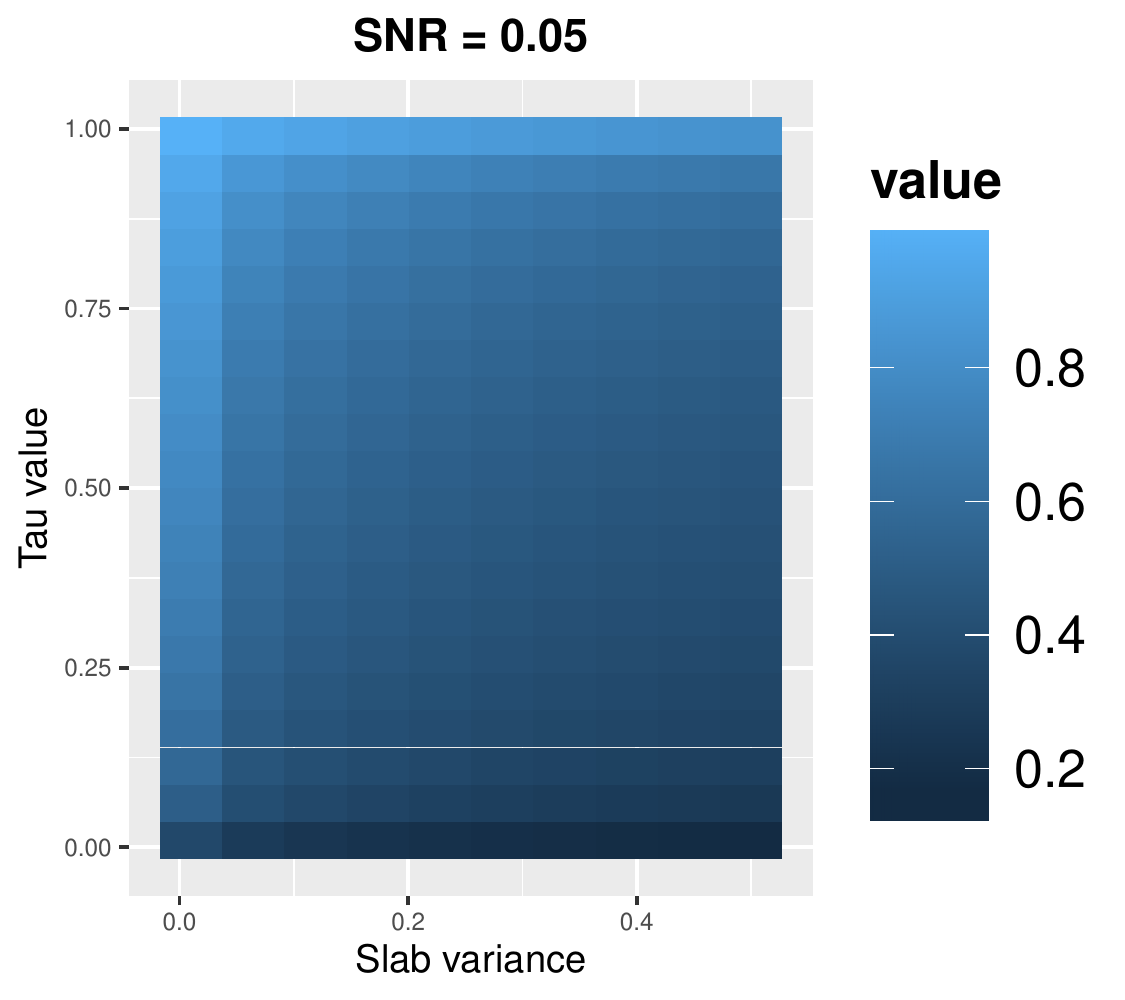}
\caption{Posterior inclusion probabilities for a range of true $\tau_{M_j}$ and $\sigma_{M}^2$ values. The left panel shows the posterior inclusion probabilities when the signal-to-noise ratio is 0, while the right panel shows the posterior inclusion probabilities when the signal-to-noise ratio is 0.05.}
\label{fig:power}
\end{figure}

There is a strong interplay between the parameters of the spike-and-slab prior and the power to detect effects. As is always the case when performing exposure selection, our goal is to maximize power while restricting the numbers of false discoveries. For ease of discussion, we focus on main effect parameters here. The same results hold for the interaction distributed lag surfaces as well and we implement the approach in simulation and data analysis for both main effects and interactions. The two key parameters in the prior distribution that govern power and false discoveries are $\tau_{M_j}$, $j=1, \ldots, p$ and $\sigma_{M}^2$. To gain insight into posterior inclusion probabilities, we examine the full conditional for $\gamma_j$, given by
\begin{align}
	P(\gamma_j \vert \boldsymbol{\theta}_{-j}, \boldsymbol{\mathcal{D}}) \propto \frac{\tau_{M_j} \phi(\boldsymbol{0}; \boldsymbol{0}, \sigma^2 \sigma_{M}^2 \boldsymbol{I})}{\phi(\boldsymbol{0}; \boldsymbol{M}_j, \boldsymbol{V}_j)} \label{eqn:Power},
\end{align}
where $\boldsymbol{\theta}_{-j}$ represents all parameters in the model except for $\gamma_j$ and $\boldsymbol{\beta}_j$. Further, $\boldsymbol{M}_j$ and $\boldsymbol{V}_j$ are the mean and variance of the full conditional distribution for $\boldsymbol{\beta}_j$ when $\gamma_j=1$.  Here $\phi(\boldsymbol{A}; \boldsymbol{B}, \boldsymbol{C})$ denotes the multivariate normal density with mean $\boldsymbol{B}$ and variance $\boldsymbol{C}$ evaluated at the vector $\boldsymbol{A}$. Clearly, if the prior inclusion probability $\tau_{M_j}$ increases, then the expression in \eqref{eqn:Power} will increase and the probability of inclusion increases for the main effect of exposure $j$. The numerator density is the prior density of the parameter vector being zero when $\gamma_j=1$, while the denominator is the density at the zero vector when using the full conditional (given data) distribution for $\boldsymbol{\beta}_j$ when $\gamma_j=1$. If the data generating mechanism truly contains a main effect, then the zero vector should be unlikely given the data, and the denominator will become  small, leading to increased posterior inclusion probabilities. The second parameter, $\sigma_M^2$, enters into both the numerator and denominator of this expression.  Therefore it is less clear how $\sigma_M^2$ relates to power, but this can be explored empirically. For a given data generating mechanism, we can calculate the average value of expression (\ref{eqn:Power}) for various values of $\sigma_M^2$, $\tau_{M_j}$, and signal-to-noise ratio. Here, the signal-to-noise ratio is calculated as $\boldsymbol{\eta}_j^T \Sigma_{x_j} \boldsymbol{\eta}_j / \sigma^2$, where $\Sigma_{x_j}$ is the covariance matrix of exposure $j$ over time and $\boldsymbol{\eta}_j$ is a $T-$dimensional vector representing the distributed lag curve. Assuming an AR(1) process for the covariance structure with correlation 0.9 between adjacent time points, Figure \ref{fig:power} shows the posterior inclusion probabilities under settings with signal-to-noise ratios of 0 and 0.05. Two things are clear from this illustration: 1) when the signal-to-noise ratio is zero, one can obtain false discoveries when $\tau_{M_j}$ is too large and/or $\sigma_M^2$ is too small and 2) when the signal-to-noise ratio is nonzero, larger values of $\tau_{M_j}$ and smaller values of $\sigma_M^2$ lead to increased power. This concept is generally true for the variance parameter $\sigma_M^2$. Smaller values of $\sigma_M^2$ lead to increased power to detect important distributed lags. Further, there is an interplay between these two prior parameters and power, which we exploit in the following section to maximize power while restricting false discoveries using the model.

\subsection{Prior elicitation strategy}
\label{sec:MaxPower}
The model above has a number of hyperparameters. We are most interested in the estimation of $\gamma_j$ and $\gamma_{jk}$ and therefore we focus attention on $\tau_{M_j}, \tau_{I_{jk}}, \sigma_{M}^2,$ and $\sigma_I^2$, as the previous section highlighted their importance to power and false discovery control. An important point is that $\tau_{M_j}$ and $\tau_{I_{jk}}$ are strictly related to the probability of inclusion, while $\sigma_{M}^2$ and $\sigma_I^2$ have an equally important purpose of shrinking coefficients corresponding to parameters that are nonzero in the model. Because of this fact, we use standard inverse-gamma prior distributions on $\sigma_{M}^2$ and $\sigma_I^2$ and estimate their posterior distributions from the data. To specify values of $\tau_{M_j}$ and $\tau_{I_{jk}}$, we take advantage of equation (\ref{eqn:Power}) in order to optimize power while limiting false discoveries. Equation (\ref{eqn:Power}) depends on $\boldsymbol{M}_j$ and $\boldsymbol{V}_j$, which take the following values as described in Supplementary Materials Section 1:
 \begin{align}
    	\boldsymbol{M}_j = \frac{1}{\sigma^2} \left(\frac{\boldsymbol{{X^*}}_j^T \boldsymbol{{X^*}}_j}{\sigma^2} + \sigma^{-2} \sigma_{M}^{-2} \boldsymbol{I} \right)^{-1} \boldsymbol{{X^*}}_j^T \boldsymbol{Y}_j^*, \ \ \ \ \boldsymbol{V}_j = \left(\frac{\boldsymbol{{X^*}}_j^T \boldsymbol{{X^*}}_j}{\sigma^2} + \sigma^{-2} \sigma_{M}^{-2} \boldsymbol{I} \right)^{-1},
\end{align}
    \noindent where we define 
    $$\boldsymbol{Y}_j^* = \boldsymbol{Y} - \boldsymbol{C} \boldsymbol{\beta}_C - \sum_{k \neq j} \sum_{t=1}^T  \boldsymbol{f}_j^T(t)\boldsymbol{\beta}_j \boldsymbol{X}_{jt} - \sum_{j_1=2}^p \sum_{j_2 < j_1} \sum_{t_1=1}^T \sum_{t_2=1}^T  \boldsymbol{f}^T_{j_1 j_2} \boldsymbol{\beta}_{j_1 j_2} (t_1, t_2) \boldsymbol{X}_{j_1 t_1} \boldsymbol{X}_{j_2 t_2},$$
    and $\boldsymbol{{X^*}}_j$ is an $n$ by $K_j$ design matrix with each column defined by $\boldsymbol{X}_{jk}^* =  \sum_{t=1}^T f_{j k}(t) \boldsymbol{X}_{jt}.$
These depend on $\boldsymbol{{X^*}}_j^T \boldsymbol{{X^*}}_j$, which is fixed, and on $\sigma^2$, $\sigma_M^2$  and $\boldsymbol{{X^*}}_j^T \boldsymbol{Y}_j^*$ which are unknown and vary as we iterate through the Markov Chain Monte Carlo (MCMC) algorithm. Our goal is to keep the marginal inclusion probability, $P(\gamma_j = 1 \vert  \boldsymbol{\mathcal{D}})$, low when exposure $j$ has no association with the outcome. If exposure $j$ has no association with the outcome, then $\boldsymbol{{X^*}}_j^T \boldsymbol{Y}_j^*$ should be centered around zero. Of course there is sampling variability around zero and we must understand this variability in order to control $P(\gamma_j = 1 \vert \boldsymbol{\mathcal{D}})$ at a particular level. Suppose our goal is to ensure that $E[P(\gamma_j = 1 \vert \boldsymbol{\mathcal{D}})] \leq \alpha$ when exposure $j$ is not associated with the outcome, where the expectation is taken with respect to the distribution of the data. Given values of $\sigma^2$ and $\sigma_M^2$, as well as the distribution of $\boldsymbol{Y}_j^*$, there is a particular value of $\tau_{M_j}$ that will ensure $E_{Y^*}[P(\gamma_j \vert \boldsymbol{\theta}_{-j},\boldsymbol{\mathcal{D}})] \leq \alpha$. Finding this value of $\tau_{M_j}$ analytically is difficult;  however, it is straightforward to find this value using a Monte Carlo approximation to this expectation. Since $Y_{ij}^*$ is approximately $N(0, \sigma^2)$ when exposure $j$ is not associated with the outcome, we can simulate what happens to $\gamma_j$ for a given set of values for $\sigma^2, \sigma_M^2$, and $\tau_{M_j}$. We draw $D$ vectors of length $n$ from a normal distribution with mean zero and variance $\sigma^2$, which we refer to as $\boldsymbol{Y}^{(d)}$ for $d=1,\dots, D$. For $d=1, \dots, D$ we draw $\gamma_j^{(d)}$ using the probability in (\ref{eqn:Power}) and keep track of $\frac{1}{D} \sum_{d=1}^D \gamma_j^{(d)}$. In particular, we find the largest value $\tau_{M_j}$ such that $\frac{1}{D} \sum_{d=1}^D \gamma_j^{(d)} \leq \alpha$. We find this value for each exposure at each iteration of the MCMC to find the value of $\tau_{M_j}$ that gives the desired posterior inclusion probability. Effectively, we are setting $\tau_{Mj} = \tau(\sigma^2, \sigma_M^2, \boldsymbol{{X^*}}_j^T \boldsymbol{{X^*}}_j)$, where the $\tau(\cdot)$ function is a deterministic function that finds the largest $\tau_{M_j}$ value for the current parameter values that gives the desired posterior inclusion probability. We can perform an analogous procedure to find $\tau_{I_{jk}}$ for each possible interaction to control the posterior inclusion probability for interactions. We demonstrate empirically in Section \ref{sec:sim} that this controls the marginal posterior inclusion probability at the desired level.

The main reason for setting $\tau_{M_j}$ and $\tau_{I_{jk}}$ in this way is to increase the posterior inclusion probabilities of important exposures and interactions. While this will lead to improved power to detect important exposures, it is conceptually different than a traditional Bayesian analysis. The parameters $\tau_{M_j}$ and $\tau_{I_{jk}}$ represent prior inclusion probabilities for the main effects and interactions, respectively, and typically represent prior beliefs in the probability of important associations. These prior parameters are then either specified {\em a priori} or are assigned hyperprior distributions that acknowledge the uncertainty in their values. We take a different approach and update these parameters throughout the MCMC to be a function of the other parameters in the model, $\sigma^2$ and $\sigma_M^2$. We demonstrate in Section \ref{sec:sim} that this approach leads to better performance in terms of exposure selection and estimation of distributed lag surfaces. 

\subsection{Identification of critical windows}

Our approach has focused on the identification of exposures and interactions that are associated with the outcome, but a secondary goal that is important in environmental epidemiology is the identification of critical windows of susceptibility. If an exposure or interaction has a critical window, it is more likely to be included in the model due to the hyperparameter selection described in Section \ref{sec:MaxPower}, but we still need an approach to understand which time periods are most associated with the outcome. We use the Bayes-$p$ measure to assess times during which there is a critical window. Formally, the Bayes-$p$ for exposure $j$ at time $t$ is defined as Bayes-$p_{jt}=1-\max\{\Pr(\eta_{jt}>0 | \mathcal{D}),\Pr(\eta_{jt}<0 | \mathcal{D})\}$.  A Bayes-$p$ near 0 indicates evidence of a critical window at that time point. We use a threshold of Bayes-$p<0.025$ to identify critical windows, which corresponds to the 95\% posterior interval not containing zero. 

\section{Simulation study}
\label{sec:sim}

Here we present the results of a simulation study to assess the performance of the proposed multiple-exposure distributed lag model.  The goals of the simulation study are to assess 1) the performance of our empirical approach to selecting basis functions on a variety of distributed lag surfaces, 2) the ability of our procedure for updating $\tau_{M_j}$ and $\tau_{I_{jk}}$ to maximize power while controlling posterior inclusion probabilities, and 3) whether our various dimension reduction strategies for interaction surfaces are able to increase power to detect weak to moderate interaction signals. To assess the first of these goals, we explore distributed lag effects that take varying functional forms and evaluate whether our basis functions are able to capture the true relationships between exposures and the outcome. To evaluate the second and third of these goals, we vary the signal-to-noise ratio by increasing the residual variance of the model and assessing the power to detect signals as a function of this residual variance. For all simulations, we vary $\sigma^2 \in \{0.5, 1, 2, 3, 4 \}$ to evaluate different signal-to-noise ratios. We explore three different variations of our model to estimate the effects of the exposures on the outcome:
\begin{enumerate}
    \item Full Bayes (FB): Our proposed model using the FPCA basis functions that retain 95\% of the variability in the exposures. This model assumes $\tau_{M_j} = \tau_M$ and $\tau_{I_{jk}} = \tau_I$ with hyperpriors for all parameters including $\tau_{M}$ and $\tau_{I}$.
    \item FB reduced: The FB model, but with dimension reduction performed on the interaction surfaces by retaining 99.9\% of the variability in the interaction basis functions. 
    \item False discovery control (FD Tau): The FB reduced model, but with $(\tau_{M_j}, \tau_{I_{jk}})$ chosen via the approach in Section \ref{sec:MaxPower}. We target posterior inclusion probabilities of 0.1 and 0.05 for null main effects and interactions, respectively. 
\end{enumerate}
We evaluate the performance of these approaches across a range of metrics including power, false discovery inclusion probability, 95\% point-wise credible interval coverage, and mean squared error (MSE). Power is defined to be the average posterior inclusion probability across all simulations, averaged over all important effects in the model. We define the false discovery inclusion probability to be the average posterior inclusion probability across all simulations, averaged over all null effects in the model. We define 95\% point-wise interval coverage to be the proportion of simulations in which the 95\% credible interval of the distributed lag curves and interaction surfaces contain the corresponding true parameters, averaged over all important parameters in the model. Lastly mean squared error is the squared difference between the posterior mean and the true values for each parameter, averaged over all exposures and time points. 

\begin{figure}[h]
\centering
	  \includegraphics[width=0.85\textwidth]{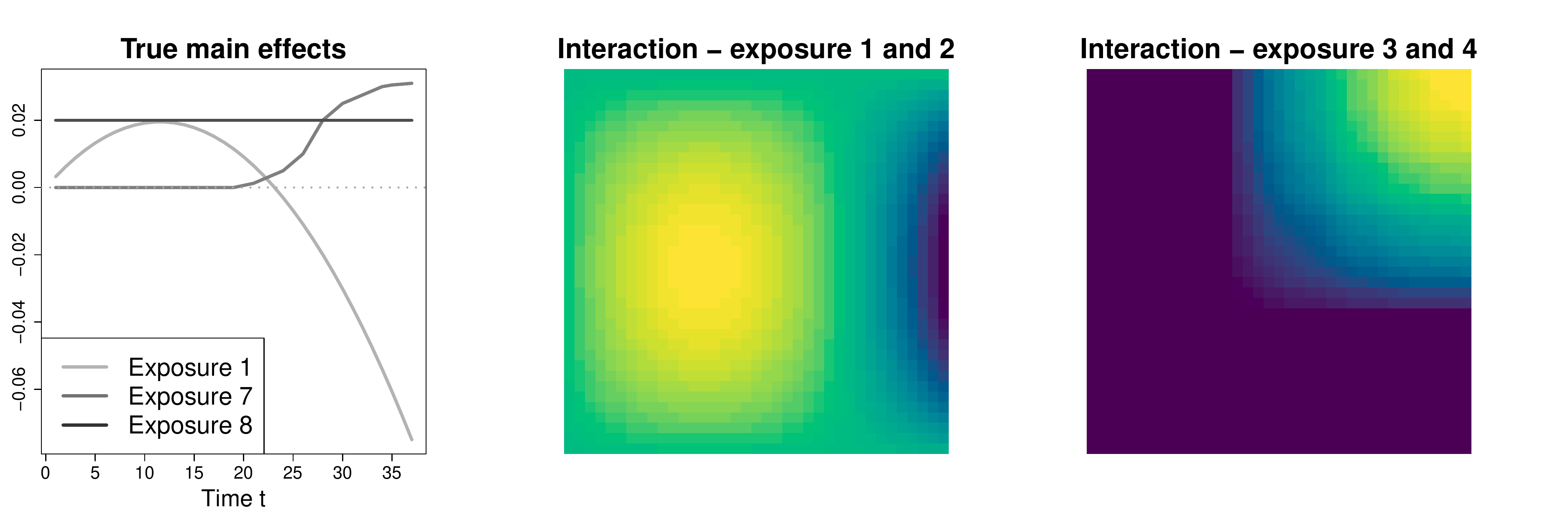} \\
\caption{The left panel shows the true main effect distributed lag curves for each exposure with an effect on the outcome, i.e. $\eta_{jt}$ for $t = 1, \dots, T$. The middle panel and the right panel show the distributed lag interaction surfaces used for the simulation study for the two pairs of interactions with effects on the outcome, i.e. $\eta_{j_1 j_2 t_1 t_2}$ for $t_1 = 1, \dots T$ and $t_2 = 1, \dots T$.}
\label{fig:TrueSurfaces}
\end{figure}

We explore one simulation scenario here and include results from additional data generating mechanisms in  Section 2 of the Supplementary Materials. Results are fairly similar across different data generating mechanisms. Throughout we set $n=400$, $p=10$ exposures in the model, and $T=37$ time points. We generate the $p$ exposures from a vector autoregressive model that allows exposures to be correlated both across exposures and within exposures across time. For each subject in the study, we draw their exposure values over time according to the following:
$$
    \boldsymbol{X}_1 = \boldsymbol{\epsilon}_1, \quad \quad
    \boldsymbol{X}_t = \boldsymbol{A} \boldsymbol{X}_{t-1} + \boldsymbol{\epsilon}_t \ \text{for } t=2, \dots, T, \quad \quad
    \boldsymbol{\epsilon}_t \sim \mathcal{MVN}(\boldsymbol{0}, \boldsymbol{\Sigma}),
$$
where $\boldsymbol{X}_t$ now denotes the $p$-dimensional vector of exposures at time $t$. We let $\boldsymbol{A}$ be a diagonal matrix with $0.95$ on the diagonals to induce temporal correlation within each exposure over time. We explore two values for the correlation among exposures within the mixture ($\boldsymbol{\Sigma}$) at a given time: independent exposures with $\Sigma_{ij} = 0$ for all $i \neq j$ and correlated exposures with $\Sigma_{ij} = 0.5^{|i -j|}$. Exposures 1, 7, and 8 have nonzero main effects and exposure pairs (1,2) and (3,4) have nonzero interaction effects. Figure \ref{fig:TrueSurfaces} shows  the functional form of the non-zero main and interaction effects. 

\subsection{Variable selection performance}
\label{sec:SimResultsInclusion}
\begin{figure}[h!]
\centering
	  \includegraphics[width=0.8\textwidth]{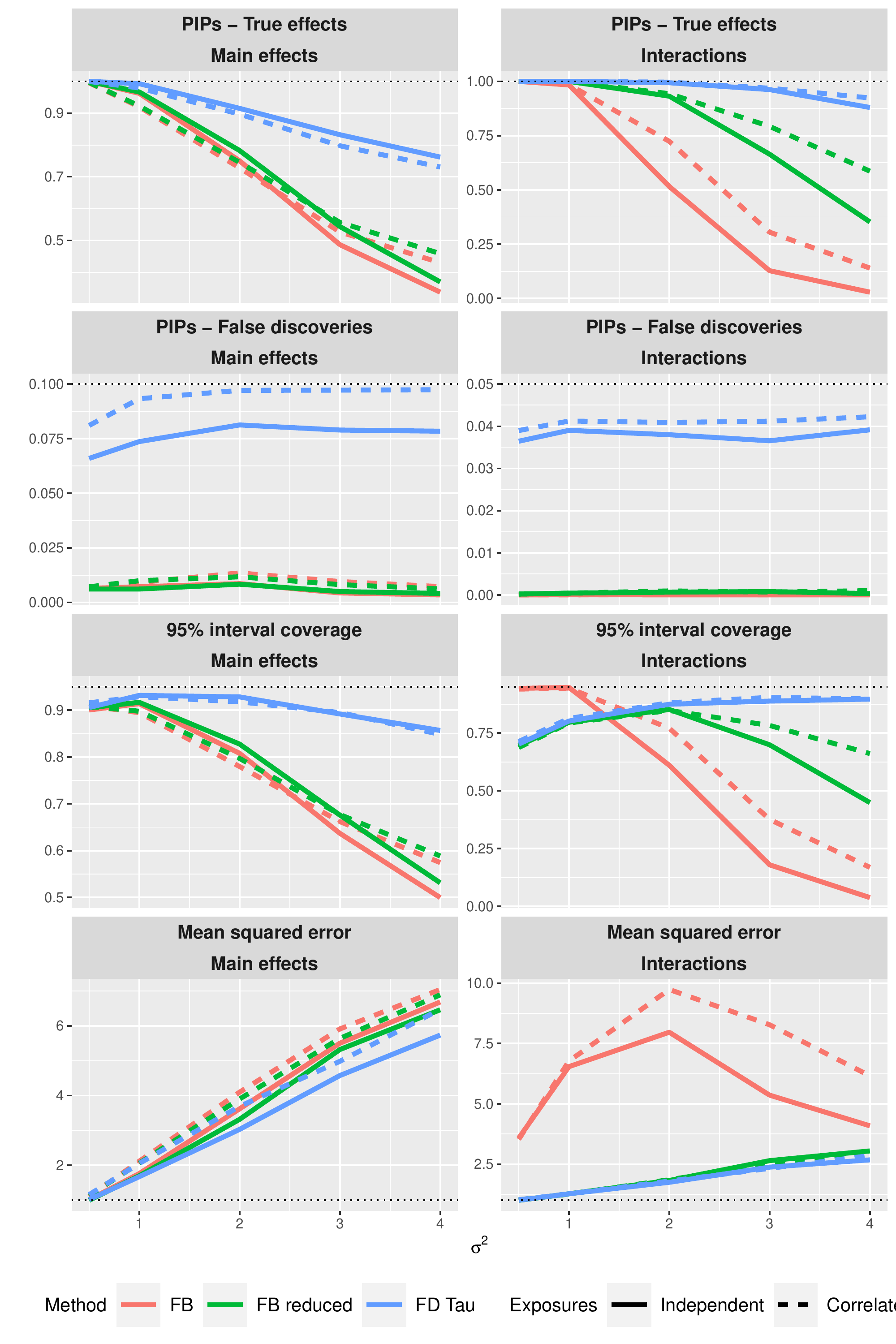}
\caption{Results from the simulation study for main effects (left column) and interactions (right column). The rows correspond to power (first row), false discoveries (second row), 95\% credible interval coverage (third row), and mean squared error (fourth row).}
\label{fig:SimResults}
\end{figure}

 Figure \ref{fig:SimResults} summarizes the simulation results. When the residual variance is 0.5, nearly all of the proposed approaches perform similarly well in terms of power, as seen in the first row of the figure. This is the setting with the greatest signal-to-noise ratio and therefore the power to detect important main and interaction effects is near 1 for all approaches. As the  residual variance increases, there is a quick decay in the power of the FB and FB reduced approaches, while the FD Tau approach maintains comparatively high power. The second row highlights the posterior inclusion probabilities for exposures or interactions that do not affect the outcome. We see these false discovery inclusion probabilities are higher for the FD Tau approach but are always below the desired levels of 0.1 for main effects and 0.05 for interactions. The performance of the approaches in both the independent and correlated exposures setting are similar.

 This increased power also leads to increased coverage rates.  Posterior inclusion probabilities are low for the true effects with the FB and FB reduced models. This leads to worse performance of the coverage of the main effects because it is less likely that resulting intervals cover the true parameter if it is not included in the model. The coverage is much closer to the nominal 95\% level for the main effects under the FD Tau approach. At low signal-to-noise ratio (higher $\sigma^2$ values), the coverage drops slightly below 0.95, although this is due to the decreasing posterior inclusion probability that occurs as the signal-to-noise ratio decreases. The MSE is lowest for the FD Tau approach in all scenarios, particularly for the interactions, which we detail in the following section. Overall, this simulation shows that the proposed approach does a good job approximating the distributed lag surfaces without a priori knowledge about their functional form.
 
\subsection{Estimation of interaction surfaces}

\begin{figure}[h]
\centering
	  \includegraphics[width=0.9\textwidth]{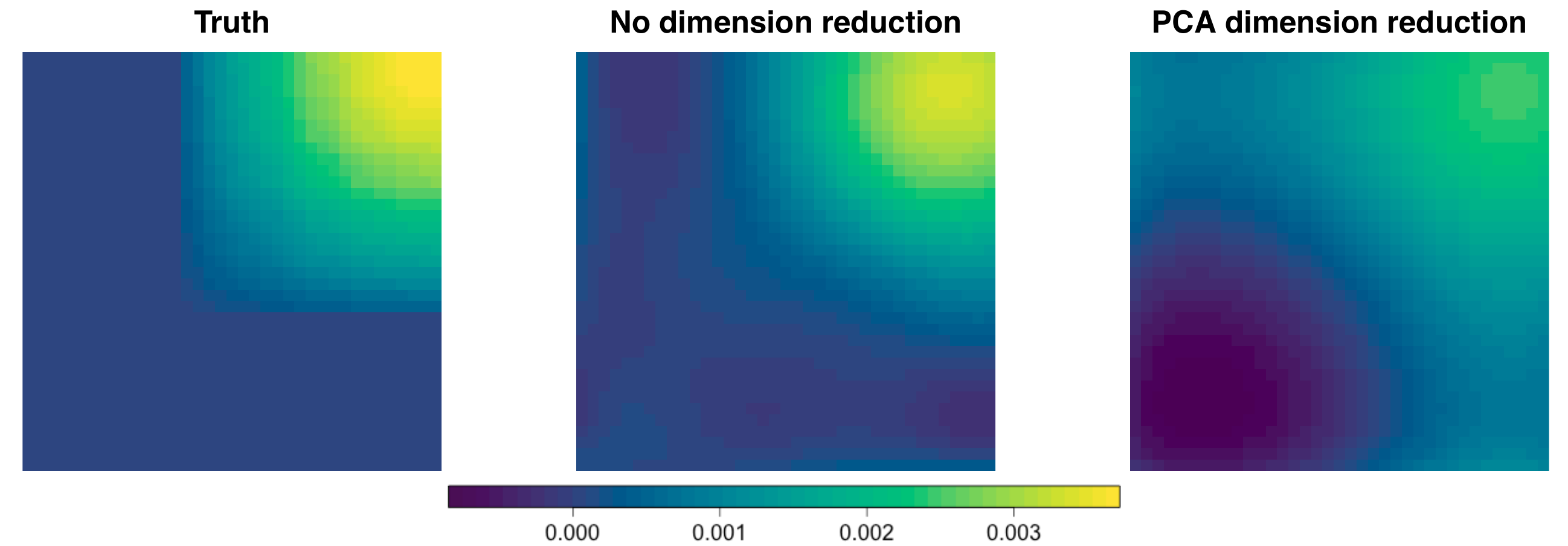}
\caption{Estimates of the interaction surface between exposures 3 and 4. The left panel shows the true surface, the middle panel shows the average estimate from the FB approach, while the right panel shows the average estimate from the FB reduced approach. }
\label{fig:IntSurfaces}
\end{figure}

All approaches have high power to detect the true interaction surfaces when $\sigma^2$ is small (high signal-to-noise ratio). The power is substantially better for the FD Tau approach for larger $\sigma^2$ values. Unlike with the main effects, the high power does not lead to 95\% credible interval coverage due to the additional dimension reduction implemented by all approaches except for the FB approach (see Section  \ref{sec:Basis}). This dimension reduction is intended to reduce the parameter space of the interaction surfaces and, as a result, leads to increased power and efficiency. We see in the top right panel of Figure \ref{fig:SimResults} that FB reduced has greater power than the FB approach. Figure \ref{fig:IntSurfaces} shows the average of the posterior means of the interaction surface for exposures 3 and 4, for both the FB and FB reduced approaches, when $\sigma^2 = 0.5$. We see that the FB model without any dimension reduction does a good job approximating the true interaction surface, but the dimension reduction approach incurs some bias in estimating this surface. This bias results from a bias-variance trade-off that leads to small amounts of bias and reduced 95\% credible interval coverage, but also improves power and efficiency of the resulting models. This efficiency gain can be seen in the bottom right panel of Figure \ref{fig:SimResults}, which shows that the MSE of the FB approach is larger than the FB reduced or FD Tau approaches that employ this dimension reduction. 

\subsection{Estimation of the cumulative exposure effects}
\label{sec:SimStandard}
A common summary parameter in distributed lag models is the cumulative effect. The cumulative effect is the expected difference in the outcome that is associated with a simultaneous one unit increase in exposure at each time point. Given that our model also includes interactions, this marginal representation of the cumulative effect requires that we fix the values of the other exposures. Throughout the rest of this work we fix the remaining exposures at their means. By zero-centering all exposures in this simulation study,  we parameterize this cumulative effect of exposure $j$ as $\delta_j=\sum_{t=1}^T \eta_{jt}$. A cumulative effect for an interaction pair can be similarly defined by fixing the remaining $p-2$ exposures. Throughout, we use the correlated exposures simulation design from above. For comparison, we consider a  model that uses average exposure over all $T$ weeks. This is a natural alternative to distributed lag models and is often used in practice. The average exposure model is nested in the distributed lag model, as they are equivalent under the constraint that $\eta_{jt_1}=\eta_{jt_2}$ for all time points $t_1$ and $t_2$.  Specifically, the average exposure model assumes that $\eta_{jt}=\delta_j/T$ for all $t=1,\dots,T$. The average exposure model is an interesting comparison model because it is popular in environmental epidemiology, and because it represents an extreme version of our proposed model with the maximum amount of dimension reduction possible. Power is now defined as the probability of significance, where an exposure is deemed significant in our model if the posterior inclusion probability is above 0.5, and an exposure is deemed significant in the averaged exposure model if it's corresponding p-value is below 0.05.

\begin{figure}[h]
\centering
	  \includegraphics[width=0.75\textwidth]{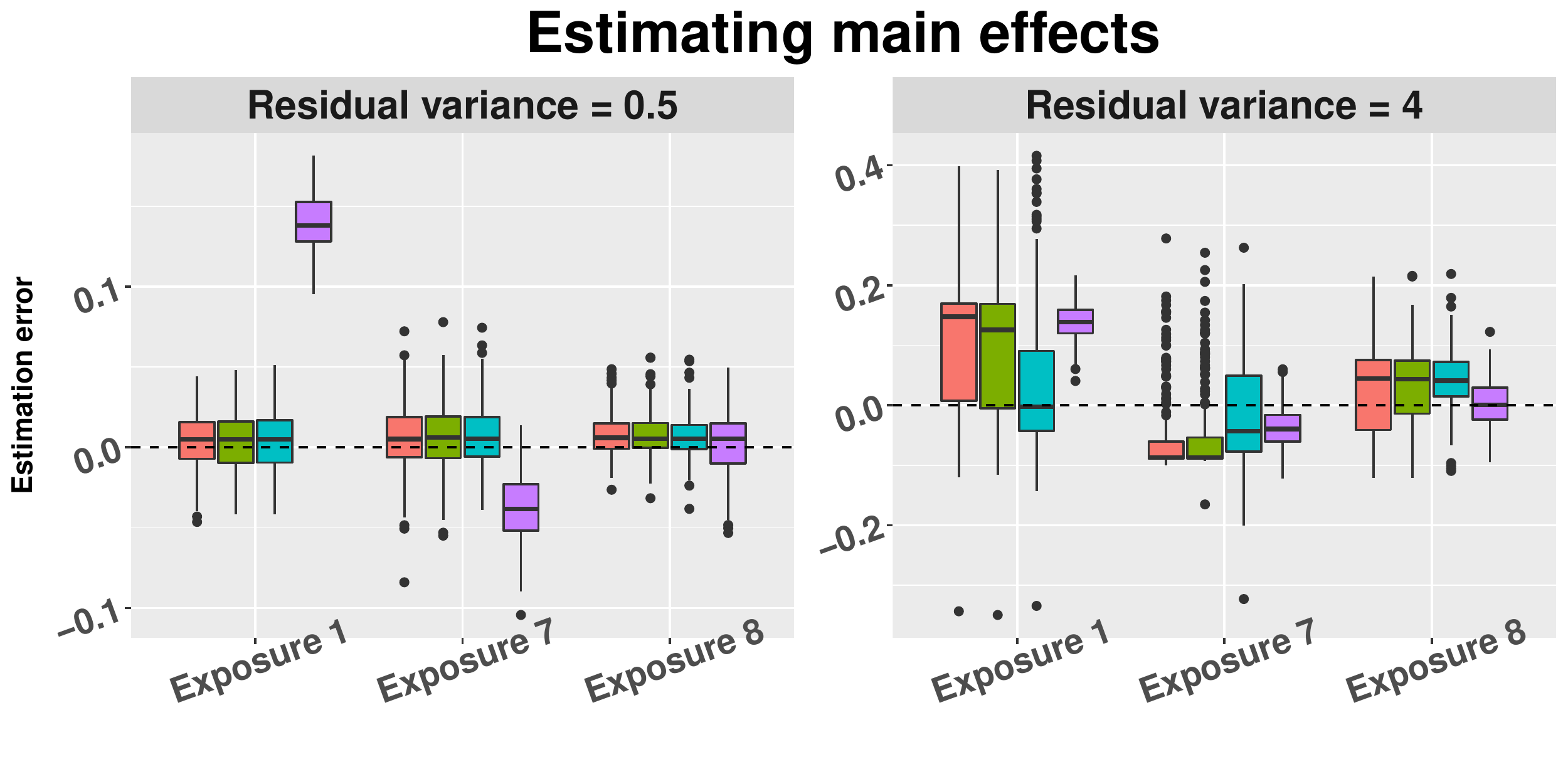} \\
	  \includegraphics[width=0.75\textwidth]{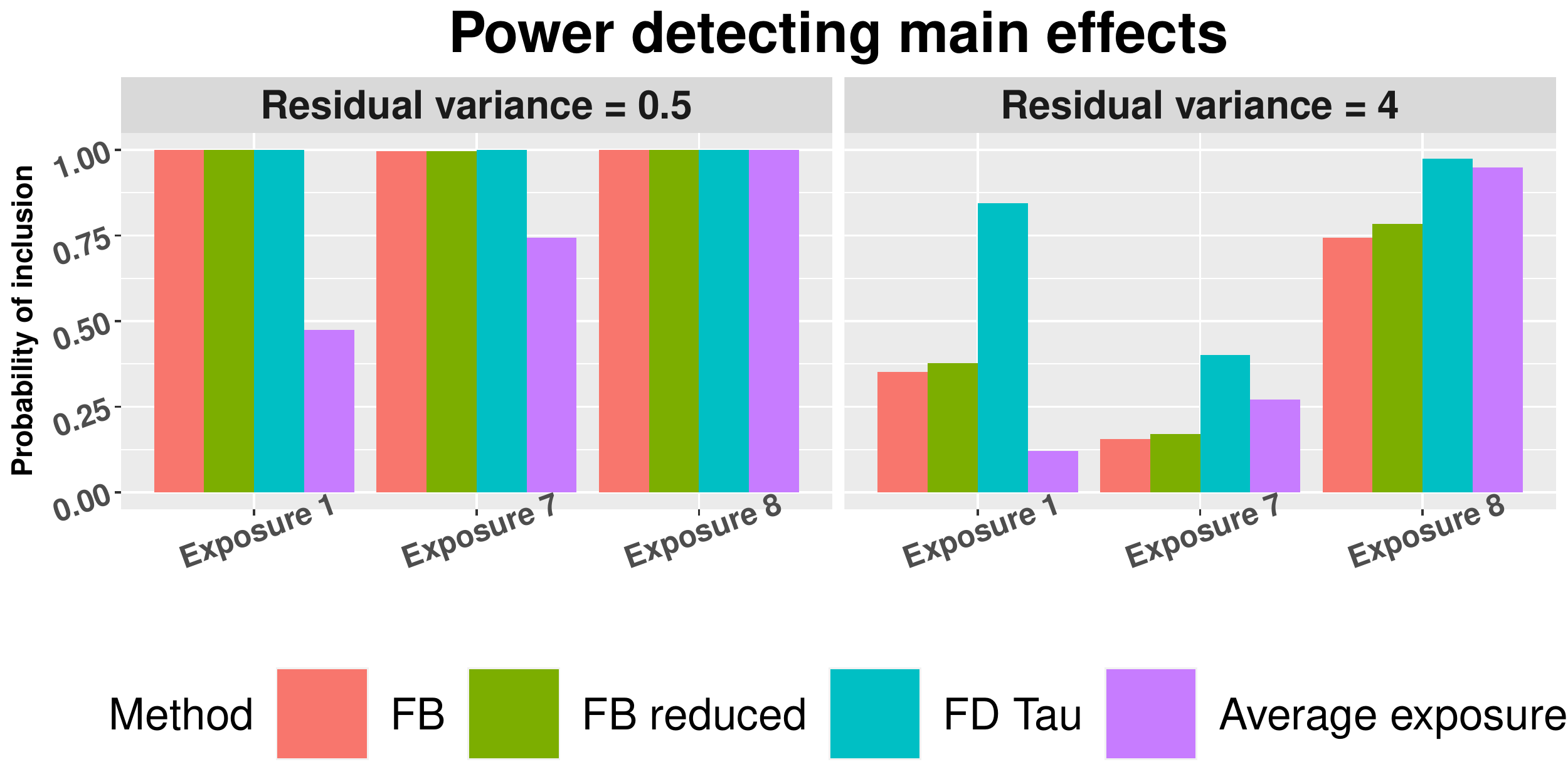}
\caption{Estimates and power for detecting the cumulative exposure effects. The top panel shows boxplots of the estimates minus their true values for distinct $\sigma^2$ values, while the bottom panel shows the power to detect each exposure.}
\label{fig:SimpleMain}
\end{figure}

Figure \ref{fig:SimpleMain} presents results for the cumulative main effect estimates.   We see that the FD Tau approach greatly outperforms the  standard approach that assigns equal weight to each time point, as well as the FB and FB reduced approaches that use distributed lag models. The estimates in the top panel of Figure \ref{fig:SimpleMain} are the estimates of the cumulative effects minus their true values. The estimators from the FD Tau approach are all nearly unbiased, while the estimates from the averaged exposure model shows substantial bias for all estimated exposures with the exception of exposure 8. Exposure 8 was assigned a flat distributed lag surface and therefore an approach that assigns equal weight to each value will perform well. As $\sigma^2$ increases, our approach suffers from increasing variability moreso than the average exposure model, but remains unbiased. The power to detect each exposure can be seen in the lower panel of Figure \ref{fig:SimpleMain}. We see that our approach has substantially higher power for each value of $\sigma^2$. The only exposure for which which the averaged exposure model has comparable power is exposure 8, which has a constant effect over time.  Even for this exposure, the power of the FD Tau approach is similar to the averaged exposure despite estimating time-varying weights, losing little in terms of power. 

\begin{figure}[h]
\centering
	  \includegraphics[width=0.75\textwidth]{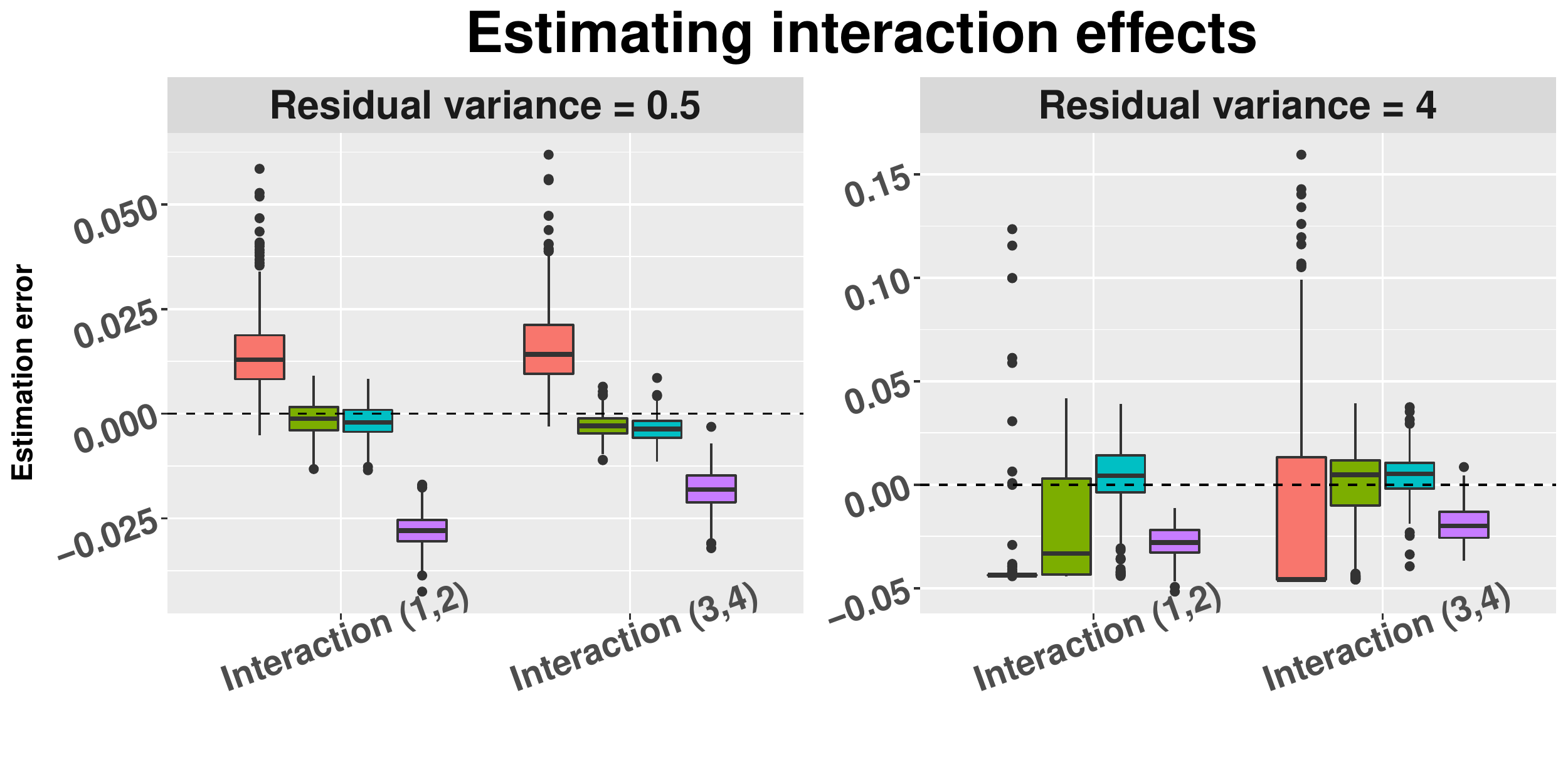} \\
	  \includegraphics[width=0.75\textwidth]{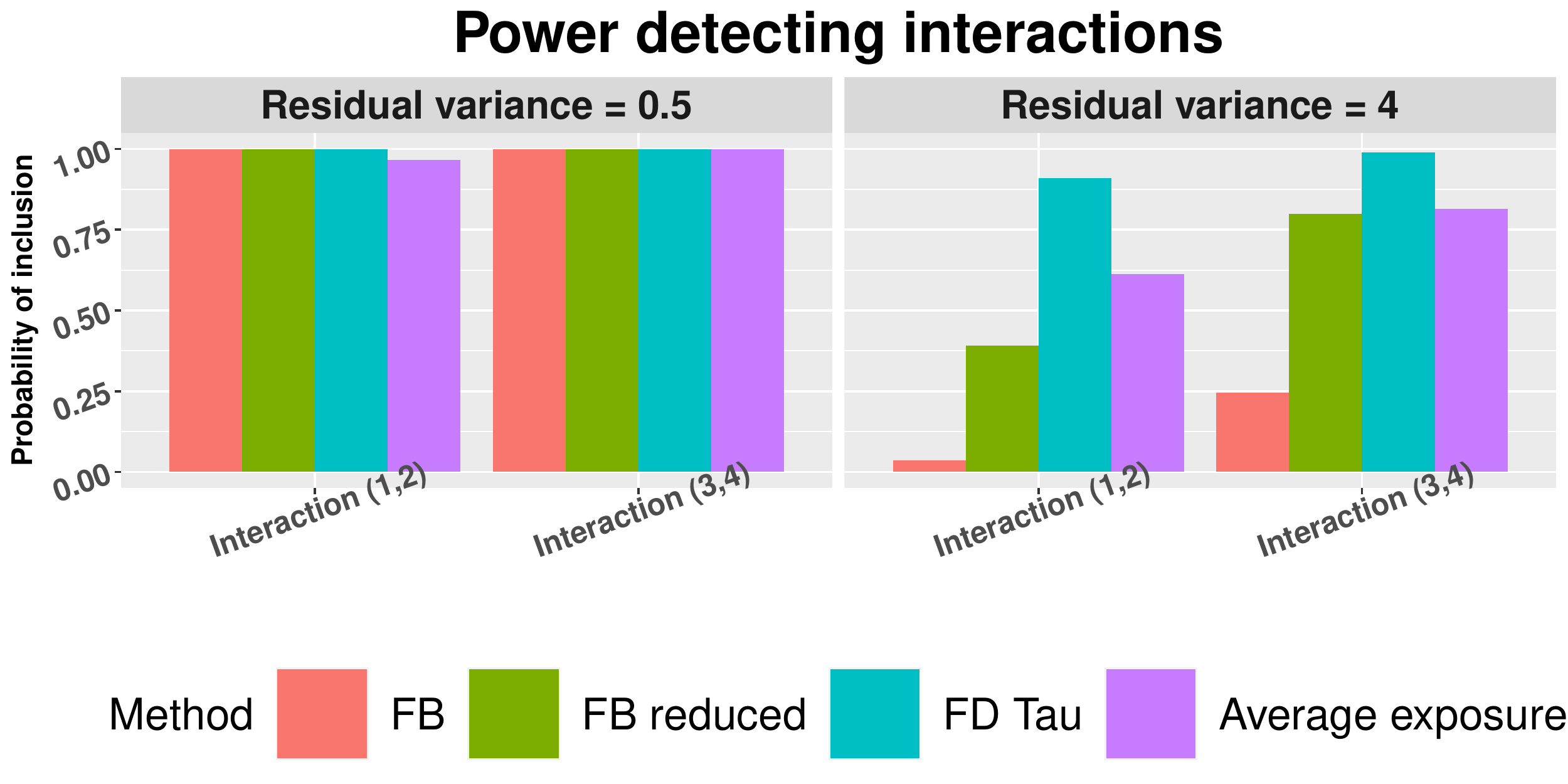}
\caption{Estimates and power for detecting the cumulative interaction effects. The top panel shows boxplots of the estimates minus their true values for distinct $\sigma^2$ values, while the bottom panel shows the power to detect each interaction.}
\label{fig:SimpleInt}
\end{figure}

Figure \ref{fig:SimpleInt}, which shows analogous results for the cumulative interaction effects, depicts a similar story. The FD Tau approach provides estimates that are closer to their corresponding true values than those from the average exposure model, which are severely biased downwards. In addition, the power to detect interactions is highest when using the FD Tau approach in all scenarios. 

\section{Analysis of Colorado data}

We apply our proposed approach in an analysis of birth weight for gestational age $z$-score (BWGAZ) and exposure to a mixture of four air pollutants and temperature. We use birth vital statistics records from Colorado, USA, with estimated conception dates between 2007 and 2015, inclusive. The five exposures of interest are particulate matter smaller than 2.5 microns in diameter (PM$_{2.5}$), nitrogen dioxide (NO$_2$), sulfur dioxide (SO$_2$), carbon monoxide (CO), and temperature. For each birth and exposure we construct  average values of estimated exposure for each week of gestation at the mother's residence.  We limit the analysis to the Denver metropolitan area for which we have more accurate exposure data. We further restrict our analysis to singleton, full-term births ($\geq37$ weeks) and observations with complete covariate and exposure data, resulting in 195,701 births. In this analysis we control for the following individual covariates: mother's age, weight, height, income, education, marital status, prenatal care habits, smoking habits, as well as race and Hispanic designations.  In addition, we include categorical variables for year and month of conception, census tract elevation, and a county-specific intercept. 
This study was approved by the Institutional Review Board of Colorado State University. 

We fit the proposed model using eight MCMC chains with 10000 iterations each, discarding the first 5000 iterations and thinning every 5th sample. We set the $\alpha$ threshold, which is the level at which we aim to control expected posterior inclusion probabilities for exposures that are not associated with the outcome, to be 0.1 for main effects and 0.05 for interactions. As a sensitivity analysis, we consider a range of $\alpha$ threshold values for the main effects and interactions. The ten candidate values are equally spaced on the log base ten scale from  0.00001 to 0.25. In the sensitivity analysis, we use the same threshold value for the main effects as the interactions. We also estimate the FB and FB reduced models that do not attempt to maximize power or control the false discovery rate, as described in Section~\ref{sec:sim}.

\subsection{Posterior inclusion probabilities}

For our primary analysis, Table~\ref{tab:PIPs} presents the PIPs. All of the main effect PIPs were  well above the $\alpha$ threshold, and all were above 0.5, which indicates an association between each exposure and birth weight.  The PIPS are  $>$0.99 for temperature, $>$0.99 for SO$_2$, 0.98 for NO$_2$, 0.98 for CO and 0.80 for PM$_{2.5}$. Four interaction surfaces had posterior inclusion probabilities above 0.5. The highest PIP values for interactions were NO$_2$-CO with a PIP of 0.93, CO-temperature with a PIP of 0.89, NO$_2$-temperature with a PIP of 0.66, and NO$_2$-SO$_2$ with a PIP of 0.60. This provides weak to moderate evidence of an interaction for these pairs of exposures. 

\begin{table}
    \centering
    \begin{tabular}{lccc}
    \hline
	&	FB	&	FB reduced	&	FD tau	\\\hline
\multicolumn{4}{l}{\textit{Main effects}}\\							
PM$_{2.5}$	&	0.01	&	0.01	&	0.80	\\
NO$_2$	&	0.01	&	0.01	&	0.98	\\
SO$_2$	&	0.00	&	0.00	&	$>$0.99	\\
CO	&	0.00	&	0.00	&	0.98	\\
Temperature	&	0.00	&	0.00	&	$>$0.99	\\	\hline
\multicolumn{4}{l}{\textit{Interaction effects}}\\						
PM$_{2.5}$ - NO$_2$	&	0.00	&	0.00	&	0.00	\\
PM$_{2.5}$ - SO$_2$	&	0.00	&	0.00	&	0.00	\\
NO$_2$ - SO$_2$	&	0.00	&	0.00	&	0.60	\\
PM$_{2.5}$ - CO	&	0.00	&	0.00	&	0.00	\\
NO$_2$ - CO	&	0.00	&	0.00	&	0.93	\\
SO$_2$ - CO	&	0.00	&	0.00	&	0.63	\\
PM$_{2.5}$ - Temperature	&	0.00	&	0.00	&	0.00	\\
NO$_2$ - Temperature	&	0.00	&	0.00	&	0.66	\\
SO$_2$ - Temperature	&	0.00	&	0.00	&	0.50	\\
CO - Temperature	&	0.00	&	0.00	&	0.89	\\\hline
    \end{tabular}
    \caption{Posterior inclusion probabilities (PIPs) for the main effects and interactions in the data analysis. The model uses an $\alpha$ threshold of 0.10 for main effects and 0.05 for interactions. The table shows results from FD Tau, which is the main model used for analysis and from the FB and FB reduced approaches which are included for sensitivity analysis.}
    \label{tab:PIPs}
\end{table}

\begin{figure}[h]
    \centering
    \subfloat[PIPs for main effects]{
     \includegraphics[width=0.85\textwidth]{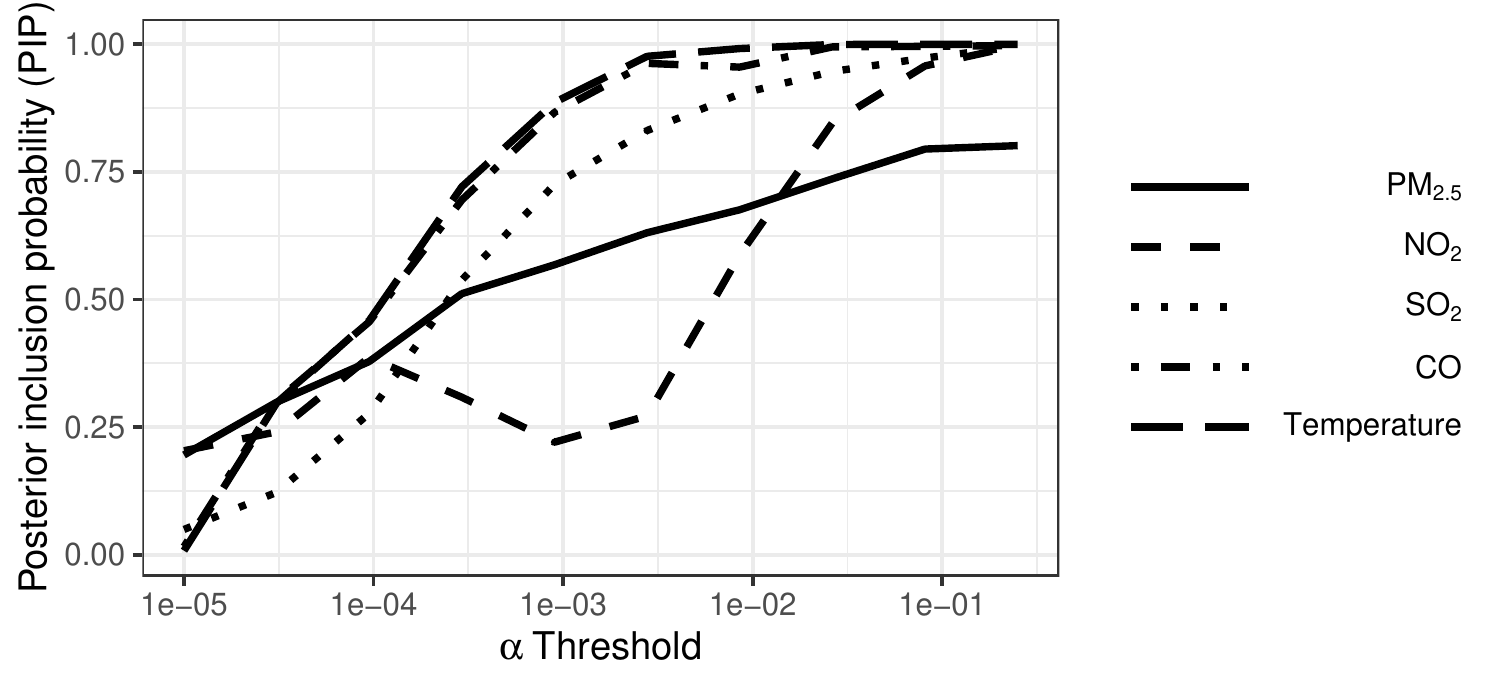}
     \label{subfig:pipmain}}
    
    \subfloat[PIPs for interaction surfaces]{
   \includegraphics[width=0.85\textwidth]{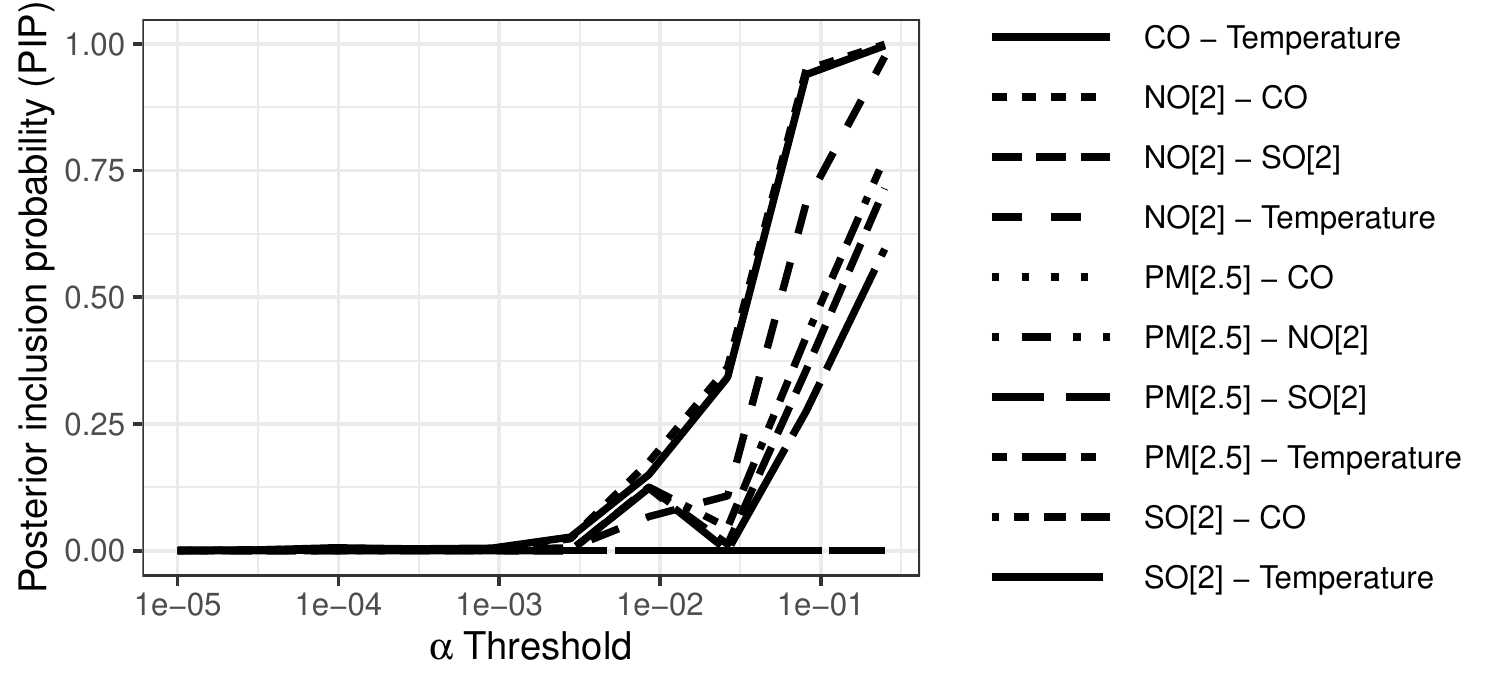}
    \label{subfig:pipint}}
    \caption{Posterior inclusion probabilities (PIPs) for the main effects (top) and interactions (bottom) of each exposure or exposure pair from the analysis of the Colorado birth cohort when we vary the $\alpha$ threshold level.}
    \label{fig:pip}
\end{figure}

Table~\ref{tab:PIPs} also shows the PIPs from FB and FB reduced. No main effects or interactions were selected by the FB or FB reduced models, which demonstrates the importance of setting hyperparameters to maximize power while controlling the false discovery rate as done by the FD Tau approach. Figure~\ref{fig:pip} shows the PIPs for the main effects and interactions for the sensitivity analysis with varying $\alpha$ thresholds. As expected, when the $\alpha$ threshold increases, the PIP for each of the component main effects and interaction surfaces increase as well. 

\subsection{Estimated distributed lag and cumulative effects}

\begin{figure}[h]
    \centering
    
    \subfloat[Estimated distributed lag functions for main effects]{
    \includegraphics[width=0.85\textwidth]{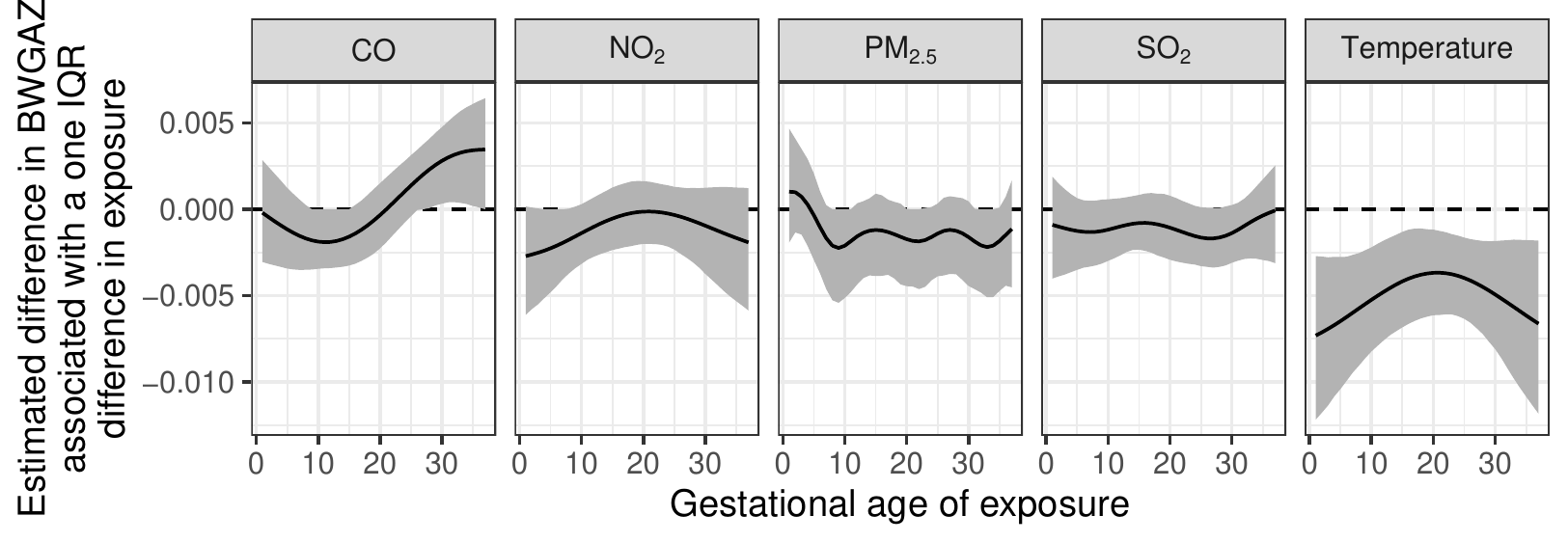}
    \label{subfig:dlm}}
    
    \subfloat[Estimated distributed lag functions for main effects]{
    \includegraphics[width=0.85\textwidth]{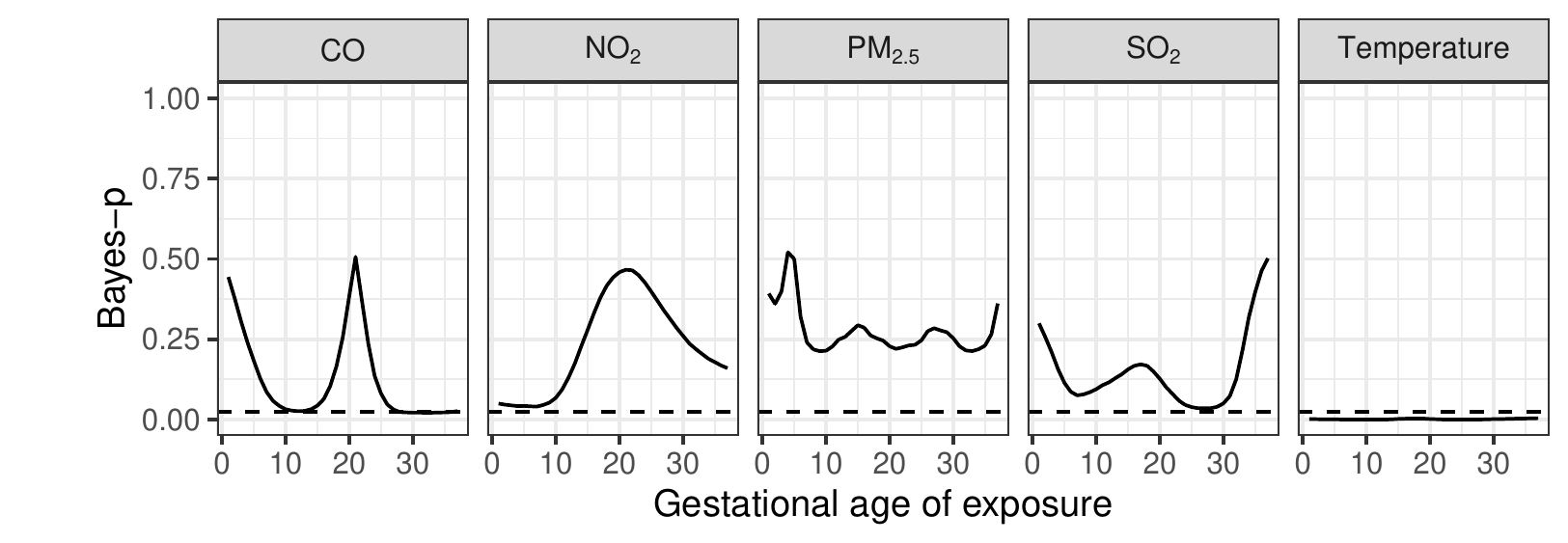}
    \label{subfig:pval}}
    
    \caption{Results of the main effect distributed lag function from the analysis of the Colorado birth cohort. Panel~\ref{subfig:dlm} shows the estimated distributed lag function for the main effect of each exposures. The figure shows the linear main effect of a one IQR exposure change on the outcome BWGAz as a function of gestational ages. Areas where the 0.95 credible interval does not cover zero are considered critical windows. Figure~\ref{subfig:pval} shows the Bayesian $p$-value for each exposure at each gestational age.} \label{fig:results}
\end{figure}

Figure~\ref{subfig:dlm} shows the posterior mean and 95\% credible interval for the main effect distributed lag functions. Figure~\ref{subfig:pval} shows the Bayes-$p$ for the association at each time point \citep{balocchi2021bayesian}. The most pronounced association was a negative association between temperature and BWGAZ. The critical window for temperature spans all 37 weeks. The estimated cumulative effect of temperature when all other exposures are fixed an their mean values is -0.19 (95\% credible interval: -0.28 to -0.10). The combination of  high PIP, clear critical windows, and a large cumulative effect that clearly departs from zero provides strong evidence of an association between temperature and BWGAZ. There was also weak evidence of an association between BWGAZ and three exposures: PM$_{2.5}$, SO$_2$ and NO$_2$.  For each of these pollutants the point estimate of the distributed lag effects was consistently negative and inferences on the cumulative effects were suggestive of an association. The estimated cumulative effect for PM$_{2.5}$ was -0.05 (credible interval: -0.11 to 0.00), for NO$_2$ was -0.04 (credible interval: -0.08 to 0.00) and for SO$_2$ was -0.04 (credible interval: -0.07 to -0.01). We did not identify a critical window for any of these components. The combination of high PIPs and suggestive or small significant cumulative effects but no critical windows provides moderately strong evidence of an association. There were two critical windows for CO and BWGAZ. A first trimester window with a negative association and a late trimester window with a positive association. However, the cumulative effect estimate was 0.01 (credible interval: -0.02 to 0.04), which suggests weak evidence of an association, particularly since the range of exposure for CO is small and the exposure is correlated with the other exposures in this data set. There was little evidence of interaction effects based on the estimated interaction surfaces as the posterior means of the interaction surfaces are concentrated around zero with intervals containing 0. For this reason we have not visualized the interaction surfaces.

\section{Discussion}

In this paper we introduced an approach to distributed lag models in the presence of multiple exposures that can identify and estimate the effects of environmental exposures over time while allowing for interactions between two exposures experienced at different points in time. We utilize spike-and-slab prior distributions to perform exposure selection for both the main effect distributed lag functions  and two-dimensional interaction surfaces, while utilizing semiparametric functions to allow the effects of exposures to vary smoothly across time. We have shown via simulation that  our approach has the ability to identify and estimate the effects of time varying exposures and analyzed the health effects of gestational exposure to air pollution on birth weight. 

One issue that makes the identification of important exposures and their interactions difficult is the lack of statistical power to detect small-to-moderate effect sizes, which are common in environmental health research.  It is a priority to maximize the power to detect these signals with limited data. We have proposed a novel approach to finding hyperparameters in our model that allows users to maximize power to detect effects while holding constant the posterior inclusion probabilities of exposures having no effect on health. Simulations demonstrated that this approach can greatly increase the posterior inclusion probabilities of important main or interaction effects. Additionally, in our analysis of the Colorado birth cohort data, we found strong evidence of an association between exposure to temperature at all weeks of gestation and decreased BWGAZ. We also saw moderate evidence of an association between the outcome and each of PM$_{2.5}$, NO$_2$, and SO$_2$, but little evidence of interactions.  This boost in power can be important for future environmental studies, many of which have a limited sample size. 

An R package implementing all approaches considered in the manuscript can be found at \url{github.com/jantonelli111/BayesianDLAG}

\section*{Acknowledgement}

This work was supported by National Institutes of Health grants ES028811, ES000002, and ES030990 and U. S. EPA grant
grant RD-835872.  These data were supplied by the Center for Health and Environmental Data Vital Statistics Program of the Colorado Department of Public Health and Environment, which specifically disclaims responsibility for any analyses, interpretations, or conclusions it has not provided. This work utilized the RMACC Summit supercomputer, which is supported by the National Science Foundation (awards ACI-1532235 and ACI-1532236), the University of Colorado Boulder and Colorado State University.  This publication's  contents are solely the responsibility of the grantee and do not necessarily represent the official views of the USEPA.

\bibliographystyle{biorefs}
\bibliography{FlexibleDLAG}

\appendix

\section{Sampling considerations}
\label{sec:sampling}

Sampling from the distributed lag model is straightforward via Gibbs sampling since all of the full conditionals have a recognizable form. Sampling proceeds from the following steps:

\begin{enumerate}
	\item When using the FB or FB reduced approaches, update $\tau_M$ from the following conditional:
    \begin{align*}
    	\tau_M \vert \bullet \sim \text{Beta} \bigg(a_{\tau_M} + \sum_{j=1}^p \gamma_j, b_{\tau_M} + p - \sum_{j=1}^p \gamma_j \bigg)
    \end{align*}
	\item When using the FB or FB reduced approaches, update $\tau_I$ from the following conditional:
    \begin{align*}
    	\tau_I \vert \bullet \sim \text{Beta} \bigg(a_{\tau_I} + \sum_{j_1=2}^p \sum_{j_2 < j_1} \gamma_{j_1 j_2}, b_{\tau_I} + \binom{p}{2} - \sum_{j_1=2}^p \sum_{j_2 < j_1} \gamma_{j_1 j_2} \bigg)
    \end{align*}
    \item For $j = 1, \dots, p$ update $\gamma_j$. If we let $\boldsymbol{\Lambda}$ represent all of the parameters in our model excluding $\boldsymbol{\beta}_j$, then our interest is in updating $\gamma_j$ from $p(\gamma_j \vert \boldsymbol{D}, \boldsymbol{\Lambda})$. We can calculate this quantity as follows:
    \begin{align*}
    	p(\gamma_j = 1 \vert \boldsymbol{D}, \boldsymbol{\Lambda}) &= \frac{p(\boldsymbol{\beta_j} = \boldsymbol{0}, \gamma_j = 1 \vert \boldsymbol{D}, \boldsymbol{\Lambda})}{p(\boldsymbol{\beta_j} = \boldsymbol{0} \vert \gamma_j = 1,  \boldsymbol{D}, \boldsymbol{\Lambda})} \\
        &\propto \frac{p(\boldsymbol{\beta_j} = \boldsymbol{0}, \gamma_j = 1)}{p(\boldsymbol{\beta_j} = \boldsymbol{0} \vert \gamma_j = 1,  \boldsymbol{D}, \boldsymbol{\Lambda})} \\
        &= \frac{\tau_{M_j} \Phi(\boldsymbol{0}; \boldsymbol{0}, \sigma^2 \sigma_{M}^2 \boldsymbol{I})}{\Phi(\boldsymbol{0}; \boldsymbol{M}_j, \boldsymbol{V}_j)}
    \end{align*}
    \noindent Using similar techniques it is easy to see that $p(\gamma_j = 0 \vert \boldsymbol{D}, \boldsymbol{\Lambda}) \propto 1 - \tau_{M_j}$, and therefore updating $\gamma_j$ just involves normalizing these probabilities and sampling from a bernoulli distribution. In this final expression, $M_j$ and $V_j$ are defined as follows:
\begin{align}
    	\boldsymbol{M}_j = \frac{1}{\sigma^2} \left(\frac{\boldsymbol{{X^*}}_j^T \boldsymbol{{X^*}}_j}{\sigma^2} + \sigma^{-2} \sigma_{M}^{-2} \boldsymbol{I} \right)^{-1} \boldsymbol{{X^*}}_j^T \boldsymbol{Y}_j^*, \ \ \ \ \boldsymbol{V}_j = \left(\frac{\boldsymbol{{X^*}}_j^T \boldsymbol{{X^*}}_j}{\sigma^2} + \sigma^{-2} \sigma_{M}^{-2} \boldsymbol{I} \right)^{-1},
\end{align}
    \noindent where we define 
    $$\boldsymbol{Y}_j^* = \boldsymbol{Y} - \boldsymbol{C}^T \boldsymbol{\beta}_C - \sum_{k \neq j} \sum_{t=1}^T  \boldsymbol{f}_j^T(t)\boldsymbol{\beta}_j \boldsymbol{X}_{jt} - \sum_{j_1=2}^p \sum_{j_2 < j_1} \sum_{t_1=1}^T \sum_{t_2=1}^T  \boldsymbol{f}^T_{j_1 j_2} \boldsymbol{\beta}_{j_1 j_2} (t_1, t_2) \boldsymbol{X}_{j_1 t_1} \boldsymbol{X}_{j_2 t_2},$$
    and $\boldsymbol{{X^*}}_j$ is an $n$ by $K_j$ design matrix with each column defined by $\boldsymbol{X}_{jk}^* =  \sum_{t=1}^T f_{j k}(t) \boldsymbol{X}_{jt}.$
    \item For $j=1,\dots,p$ update $\boldsymbol{\beta}_j$. If $\gamma_j = 0$, then $\boldsymbol{\beta}_j = \boldsymbol{0}$, otherwise update $\boldsymbol{\beta}_j$ from a multivariate normal distribution with mean $\boldsymbol{M}_j$ and variance $\boldsymbol{V}_j$. 
    \item For all combinations of $j_1$ and $j_2$ update $\gamma_{j_1 j_2}$. If we let $\boldsymbol{\Lambda}$ represent all of the parameters in our model excluding $\boldsymbol{\beta}_{j_1 j_2}$, then we can perform a similar sampling technique as for $\gamma_j$ by calculating
        \begin{align*}
    	p(\gamma_{j_1 j_2} = 1 \vert \boldsymbol{D}, \boldsymbol{\Lambda}) \propto \frac{\tau_{I_{j_1 j_2}} \Phi(\boldsymbol{0}; \boldsymbol{0}, \sigma^2 \sigma_{I}^2 \boldsymbol{I})}{\Phi(\boldsymbol{0}; \boldsymbol{M}_{j_1 j_2}, \boldsymbol{V}_{j_1 j_2})}.
    \end{align*}
    Similarly to before, $M$ and $V$ are defined as follows:
        \begin{align*}
    	\boldsymbol{M}_{j_1 j_2} = \frac{1}{\sigma^2} \left(\frac{\boldsymbol{{X^*}}^T \boldsymbol{{X^*}}}{\sigma^2} + \sigma^{-2} \sigma_{I}^{-2} \boldsymbol{I} \right)^{-1} {\boldsymbol{{X^*}}}^T \boldsymbol{Y}_{j_1 j_2}^*, \ \ \ \ \boldsymbol{V}_{j_1 j_2} = \left(\frac{\boldsymbol{{X^*}}^T \boldsymbol{{X^*}}}{\sigma^2} + \sigma^{-2} \sigma_{I}^{-2} \boldsymbol{I} \right)^{-1}.
    \end{align*}
    \noindent Where we define $$\boldsymbol{Y}_{j_1 j_2}^* = \boldsymbol{Y} - \boldsymbol{C}^T \boldsymbol{\beta}_C - \sum_{j=1}^p \sum_{t=1}^T \boldsymbol{f}^T_j(t) \boldsymbol{\beta}_j  \boldsymbol{X}_{jt} - \sum_{j, k: (j, k) \neq (j_1, j_2)} \sum_{t_1=1}^T \sum_{t_2=1}^T \boldsymbol{f}_{j k}^T (t_1, t_2) \boldsymbol{\beta}_{j k} \boldsymbol{X}_{j t_1} \boldsymbol{X}_{k t_2}$$
    and $\boldsymbol{X}^*$ is an $n$ by $K_{j_1} K_{j_2}$ design matrix with elements defined by $$ \sum_{t_1=1}^T \sum_{t_2=1}^T f_{j_1 k_{j_1}}(t_1) f_{j_2 k_{j_2}}(t_2) X_{j_1 t_1 i} X_{j_2 t_2 i},$$
    and each $(k_{j_1}, k_{j_2})$ pair corresponds to one of the $K_{j_1} K_{j_2}$ columns of the design matrix. 
    \item For each combination of $j_1$ and $j_2$, set $\boldsymbol{\beta}_{j_1 j_2} = \boldsymbol{0}$ if $\gamma_{j_1 j_2} = 0$, otherwise update it from a multivariate normal distribution with mean $\boldsymbol{M}_{j_1 j_2}$ and variance $\boldsymbol{V}_{j_1 j_2}$
    \item Update $\sigma^2$ from the following full conditional:
    \begin{align*}
    	\sigma^2 \vert \bullet \sim \text{InvGamma} \Bigg( a_{\sigma^2} + n/2 + \sum_{j=1}^p \gamma_j K_j/2 + \sum_{j_1=2}^p \sum_{j_2 < j_1} \gamma_{j_1 j_2} K_{j_1} K_{j_2}/2, \\
        b_{\sigma^2} + \text{SS}/2 + \sum_{j=1}^p \frac{|| \boldsymbol{\beta}_j ||_2^2}{2 \sigma_{M}^2} + \sum_{j_1=2}^p \sum_{j_2 < j_1} \frac{|| \boldsymbol{\beta}_{j_1 j_2} ||_2^2}{2 \sigma_{I}^2} \Bigg),
    \end{align*}
    \noindent where $$\text{SS} = \sum_{i=1}^n \Big(Y_i - \boldsymbol{C}_i^T \boldsymbol{\beta}_C - \sum_{j=1}^p \sum_{t=1}^T \boldsymbol{f}^T_j(t) \boldsymbol{\beta}_j  X_{jti} - \sum_{j_1=2}^p \sum_{j_2 < j_1} \sum_{t_1=1}^T \sum_{t_2=1}^T \boldsymbol{f}^T_{j k} (t_1, t_2) \boldsymbol{\beta}_{j k}  X_{j_1 t_1 i} X_{j_2 t_2 i}\Big)^2$$
\end{enumerate}

\section{Additional simulation studies}

Here we run additional simulation studies with different data generating mechanisms from the main manuscript. Specifically we look at simulations that involve different distributed lag curves, and simulations where there are no exposure effects.  

\subsection{No exposure effects}

\begin{figure}
\centering
	  \includegraphics[width=0.95\textwidth]{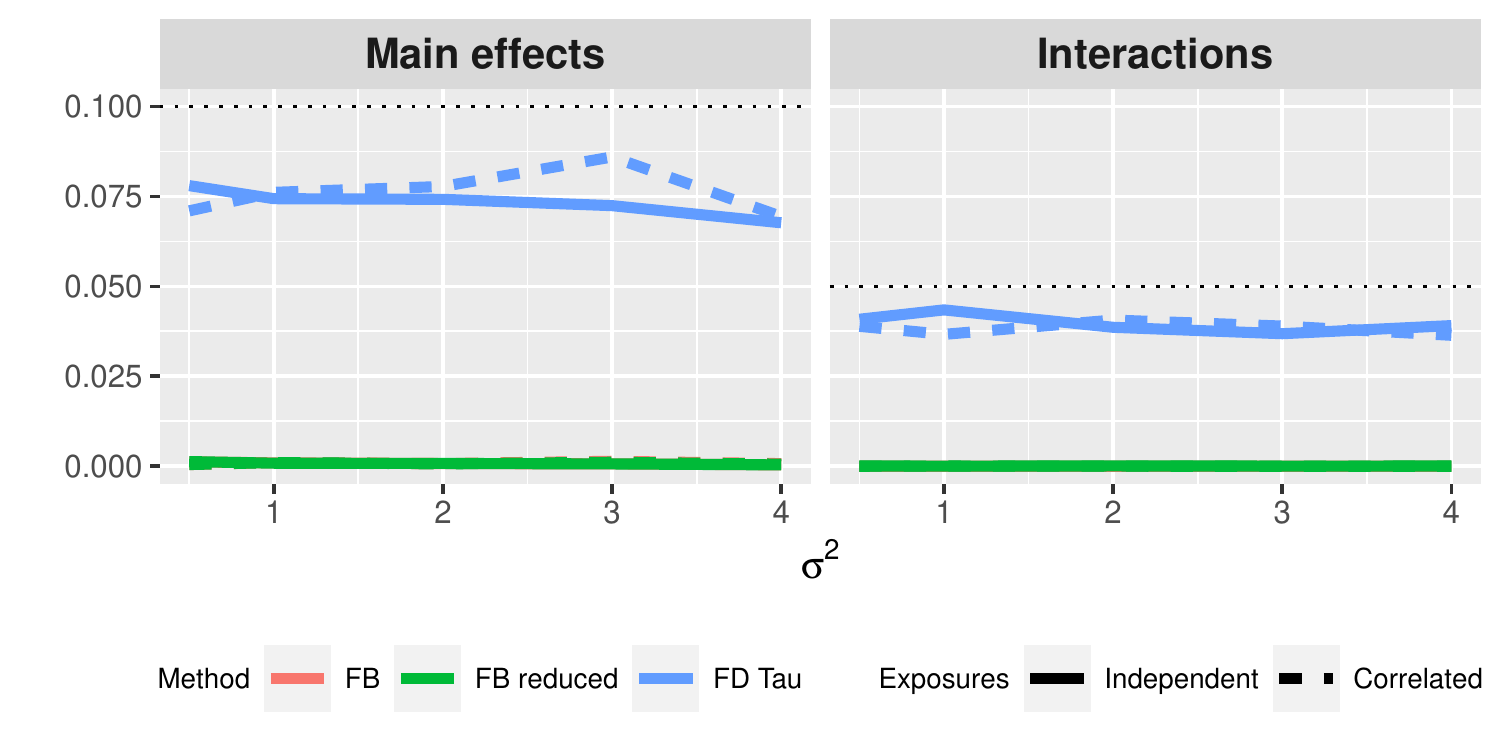}
\caption{Posterior inclusion probabilities for main effect and interaction distributed lags as a function of the residual error in the simulation with no exposure or interaction effects. The dotted lines represent the desired level of the FD Tau approach.}
\label{fig:AppendixNullSim}
\end{figure}

In this simulation study, we again let $p=10$ and we generate our exposures from a multivariate normal distribution as described in the manuscript. We now fix $n=200$ and simulate data in which there is no association between the exposures and outcomes. For this reason, we focus attention on results involving the posterior inclusion probabilities for the exposures and interactions. Specifically, our approach is designed to limit these to 0.1 and 0.05 for main effects and interactions, respectively. The results can be found in Figure \ref{fig:AppendixNullSim}, and we see that the FD Tau approach is doing as expected by obtaining posterior inclusion probabilities near 0.1 for main effects and 0.05 for interactions. The FB and FB reduced approaches have smaller posterior inclusion probabilities that are very close to zero in this setting, similar to what was seen in the simulations of the manuscript.  

\subsection{Main effects only}

\begin{figure}[h]
\centering
	  \includegraphics[width=0.45\textwidth]{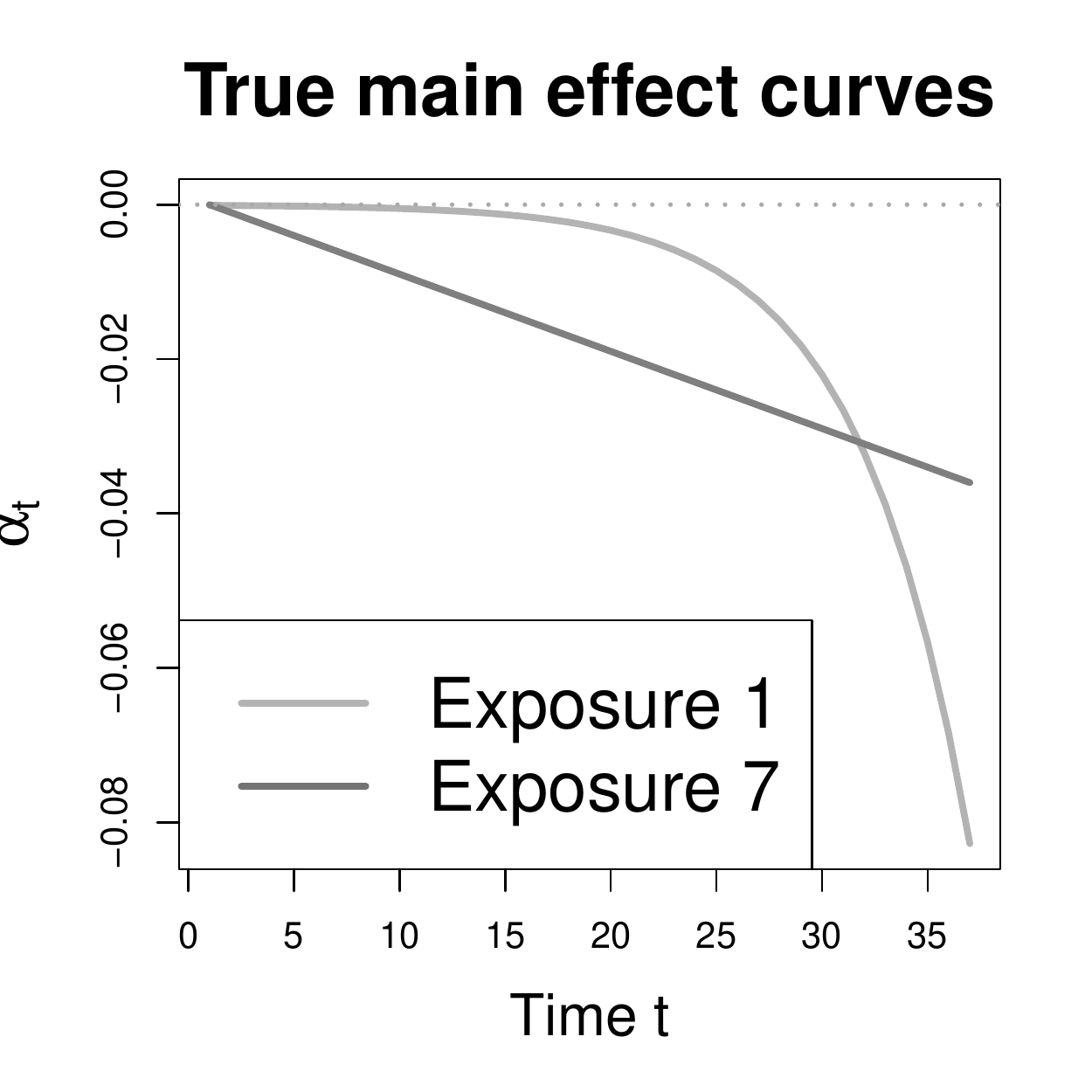}
\caption{True distributed lag curves for exposures 1 and 7 in the simulation with main effects only. }
\label{fig:AppendixMainSimTrue}
\end{figure}

Now we simulate data with main effects only and no interactions between the exposures. We again draw the exposures from the same multivariate distribution used in the manuscript and we set $n = 300$ in this setting. The two exposures with important main effects are exposures 1 and 7, and their distributed lag curves can be seen in Figure \ref{fig:AppendixMainSimTrue}. 

\begin{figure}
\centering
\includegraphics[width=0.95\textwidth]{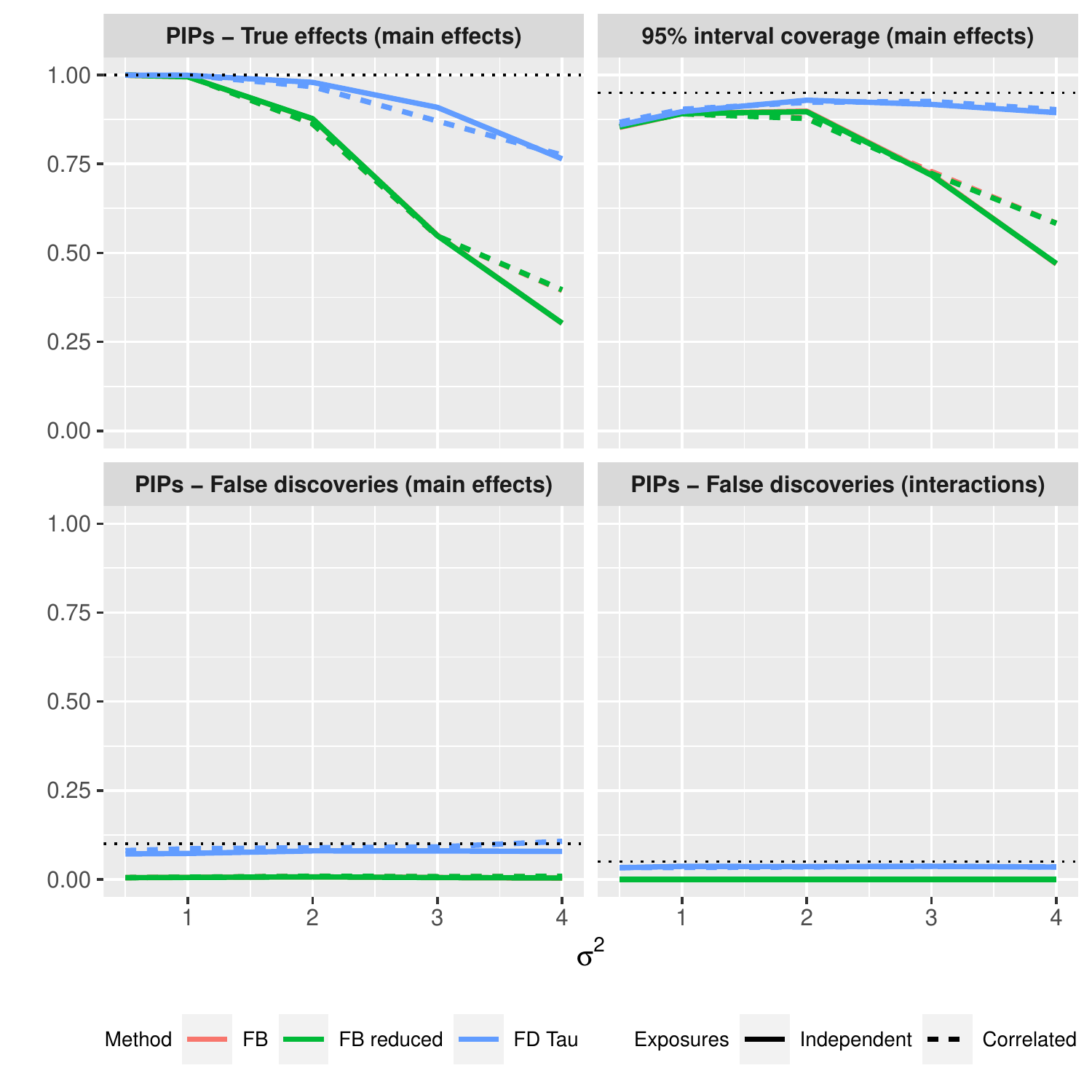}
\caption{Results of the simulation study with main effects only. }
\label{fig:AppendixMainSimResults}
\end{figure}

\begin{figure}
\centering
    \includegraphics[width=0.95\textwidth]{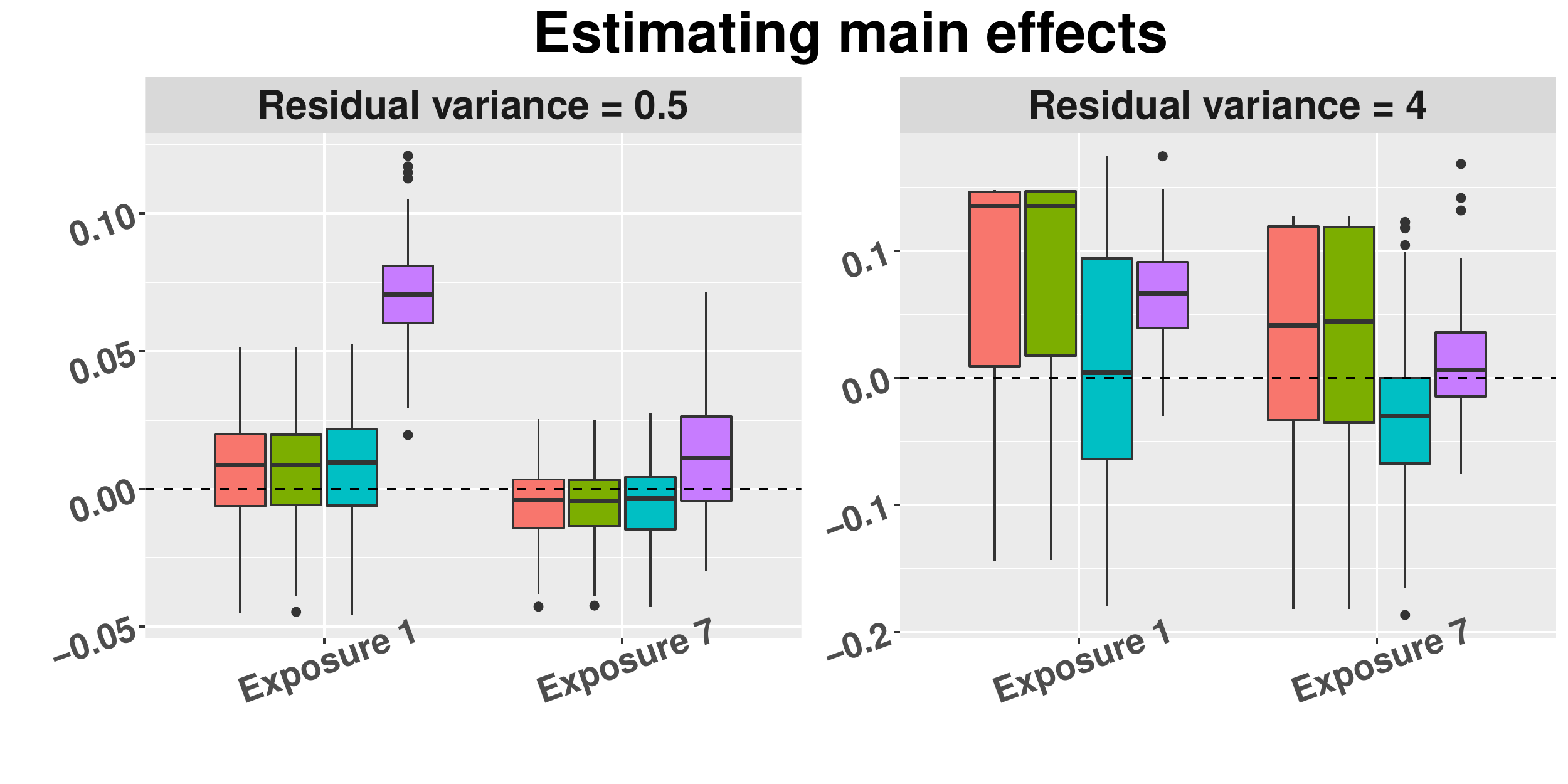}
	  \includegraphics[width=0.95\textwidth]{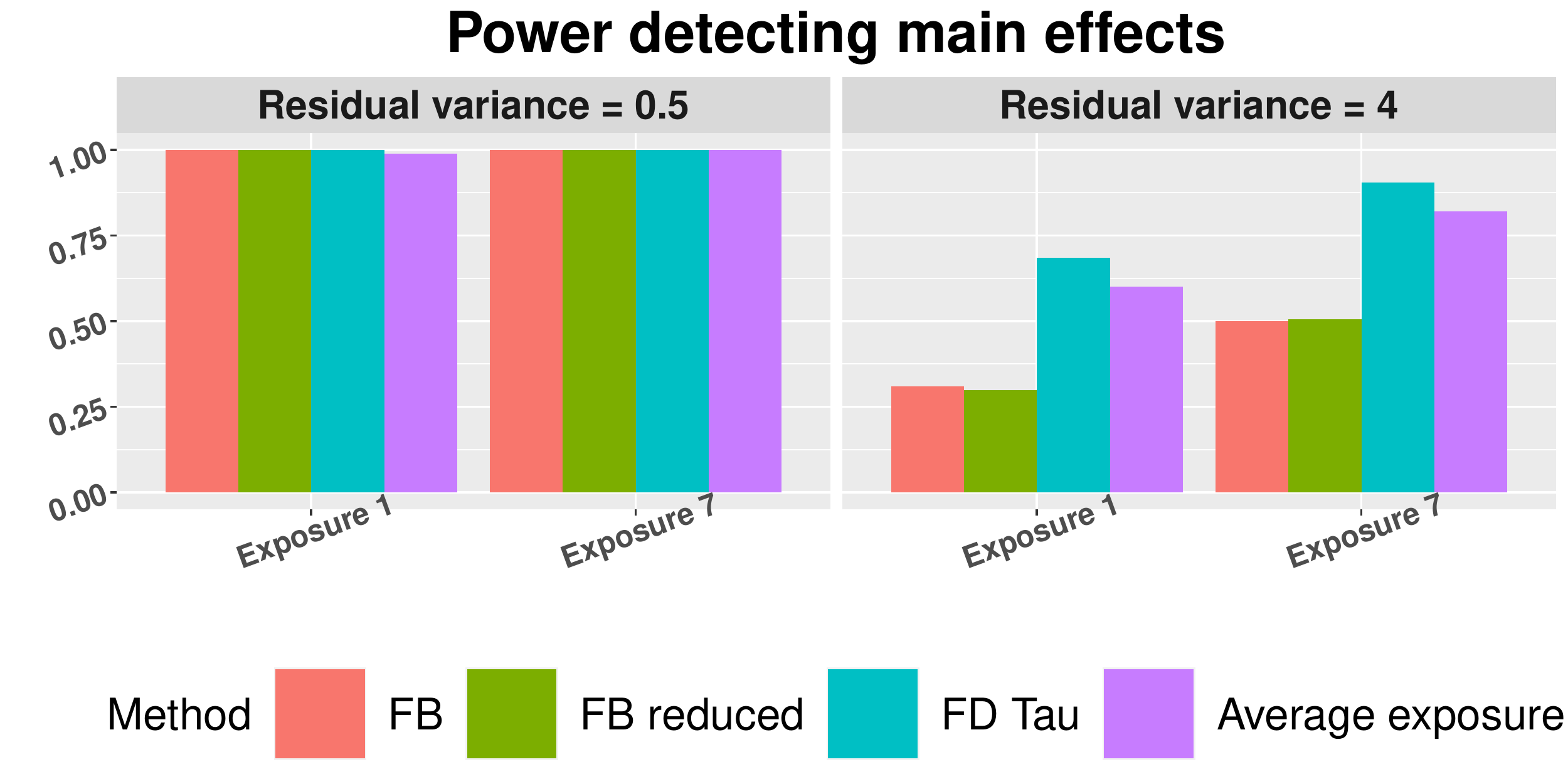}
\caption{Results of the simulation study with main effects only that focuses on the cumulative exposure effect for the two exposures with important distributed lags.}
\label{fig:AppendixMainSimSimple}
\end{figure}

The results from this simulation study can be found in Figures \ref{fig:AppendixMainSimResults} and \ref{fig:AppendixMainSimSimple}. Overall the results mirror those seen in the main manuscript. The basis functions again perform well as the interval coverage for the important main effects are close to the nominal 95\% level. The FD Tau approach again has increased power relative to the FB and FB reduced approaches. The FD Tau approach has posterior inclusion probabilities for the null main and interaction effects that is roughly equal to the 0.1 and 0.05 levels that it is targeting. Figure \ref{fig:AppendixMainSimSimple} shows that the averaged exposure approach leads to biased estimates of the overall exposure effects, while the FD Tau approach leads to estimates with low amounts of bias and increased power relative to the FB, FB reduced, and averaged exposure approaches. 

\end{document}